%% file: main.tex
\documentclass[sigconf]{acmart}

\PassOptionsToPackage{table}{xcolor}
\usepackage{pifont}
\usepackage{fancyvrb}
\usepackage{fvextra}
\usepackage{subcaption}
\usepackage{pdfrender}
\newcommand*{\boldcheckmark}{
  \textpdfrender{
    TextRenderingMode=FillStroke,
    LineWidth=1.5pt,
  }{\checkmark}
}
    
\setcopyright{cc}
\setcctype{by}

\input{sections/defs}

\setcopyright{acmlicensed}

\copyrightyear{2026}
\acmYear{2026}
\setcopyright{cc}
\setcctype{by}

\renewcommand\footnotetextcopyrightpermission[1]{}
\acmConference[CCS '26]{Proceedings of the 2026 ACM SIGSAC Conference on Computer and Communications Security}{November 15--19, 2026}{The Hague, The Netherlands}

\acmDOI{XXXXXXX.XXXXXXX}
\acmISBN{}

\begin{document}
\title{Identifying AI Web Scrapers Using Canary Tokens}
\author{Steven Seiden}
\affiliation{
  \institution{Duke University}
  \department{Department of Electrical and Computer Engineering}
  \city{Durham}
  \state{NC}
  \country{USA}
}

\author{Triss Ren}
\affiliation{
  \institution{University of Pittsburgh}
  \department{Department of Computational and Systems Biology}
  \city{Pittsburgh}
  \state{PA}
  \country{USA}
}

\author{Caroline Zhang}
\affiliation{
  \institution{Duke University}
  \department{Department of Computer Science}
  \city{Durham}
  \state{NC}
  \country{USA}
}

\author{Taein Kim}
\affiliation{
  \institution{Duke University}
  \department{Department of Electrical and Computer Engineering}
  \city{Durham}
  \state{NC}
  \country{USA}
}

\author{Enze Liu}
\orcid{0000-0003-4288-8485}
\affiliation{
  \institution{Carnegie Mellon University}
  \department{CyLab Security \& Privacy Institute}
  \city{Pittsburgh}
  \state{PA}
  \country{USA}
}

\author{Emily Wenger}
\affiliation{
  \institution{Duke University}
  \department{Department of Electrical and Computer Engineering}
  \city{Durham}
  \state{NC}
  \country{USA}
}

\input{abs}


\maketitle
\newtcolorbox{callout}{
colback=gingerbeer!80,
colframe=black,
coltext=black,
boxrule=1pt,
arc=5pt,
left=3pt,
right=3pt,
top=2pt,
bottom=2pt,
}

\newcommand{\cmark}{\ding{51}}
\newcommand{\xmark}{\ding{55}}

\input{acronyms}
\input{sections/1_Intro}
\input{sections/2_Background}
\input{sections/3_SystemDesign}

\input{sections/4_ExpSetup}
\input{sections/5_Results}
\input{sections/6_Discussion}

\input{sections/7_Conclusion}
\input{sections/ack}

\bibliographystyle{ACM-Reference-Format}
\bibliography{MyCite}

\appendix
\input{sections/appendix_a}

\input{sections/appendix_b}
\input{sections/appendix_c}

\input{sections/appendix_d}

\input{sections/appendix_e}

\end{document}

%% file: sections/defs.tex
\usepackage{booktabs}
\usepackage[nolist]{acronym}
\usepackage{caption}
\usepackage{tikz}
\usepackage{multirow, multicol}
\usepackage{float}
\usetikzlibrary{shapes.geometric, arrows.meta, positioning}
\usepackage{float}
\usepackage{makecell}
\usepackage{xcolor}
\usepackage{longtable}
\usepackage{float}
\usepackage{enumitem}

\newcommand{\para}[1]{{\vspace{2pt} \bf \noindent #1}} 

\tikzstyle{startstop} = [rectangle, rounded corners, minimum width=3cm, minimum height=1cm,text centered, draw=black, fill=gray!20]
\tikzstyle{decision} = [diamond, draw, fill=blue!10, text centered, inner sep=1pt, aspect=2]
\tikzstyle{process} = [rectangle, minimum width=3cm, minimum height=1cm, text centered, draw=black, fill=green!20]
\tikzstyle{arrow} = [thick,->,>=stealth]

\definecolor{copper}{HTML}{C84E00}
\definecolor{dandelion}{HTML}{FFD960}
\definecolor{piedmont}{HTML}{A1B70D}
\definecolor{eno}{HTML}{339898}
\definecolor{shale}{HTML}{0577B1}
\definecolor{dukeblue}{HTML}{012169}
\definecolor{ironweed}{HTML}{993399}
\definecolor{whisper}{HTML}{F3F2F1}
\definecolor{gingerbeer}{HTML}{FCF7E5}
\hypersetup{
  colorlinks=true,
  linkcolor=dukeblue!70!white,
  urlcolor=dukeblue!70!white,
  citecolor=dukeblue!70!white
}

\newcommand{\todocolorlist}[1]{%
  \ifcase#1\or red\or cyan\or orange\or violet\or green\or magenta\or purple\or blue\or brown\or olive\or teal\or lime\fi%
}
\newcommand{\todolistlength}{5}

\newcounter{todocoloridx}
\newcommand{\MakeTodo}[2][]{%
  \ifx&#1&%
    \colorlet{todocolor#2}{\todocolorlist{\thetodocoloridx}}%
    \stepcounter{todocoloridx}%
    \ifnum\thetodocoloridx>\todolistlength%
      \setcounter{todocoloridx}{1}%
    \fi%
  \else
    \colorlet{todocolor#2}{#1}%
  \fi
  \expandafter\let\csname #2\endcsname\undefined%
  \expandafter\newcommand\csname #2\endcsname[1]{%
    \textcolor{todocolor#2}{\textbf{\ [~{#2}:~##1~]}}%
  }%
}

\newcommand{\rowheight}{\rule{0pt}{3.2ex}}

\usepackage[most]{tcolorbox}

\newtcolorbox{prettyquote}[2][]{%
    enhanced,
    breakable,
    colback=gray!5,
    colframe=gray!40,
    boxrule=0.4pt,
    arc=1.5mm,
    left=2mm,
    right=2mm,
    top=1.5mm,
    bottom=1.5mm,
    width=\linewidth,
    after={\par\vspace{0.3em}\centering\small\refstepcounter{figure}\textbf{Figure \thefigure:} #2\par\vspace{0.5em}},
    #1
}

%% file: abs.tex
\begin{abstract}

From pre-training to query-time augmentation, web-scraped data helps to improve the quality and contextual relevancy of content generated by AI chatbots such as ChatGPT and Claude. However, large-scale web scraping to feed such chatbots can affect site stability and raise legal, privacy, or ethics concerns. If website owners wish to limit AI-chatbot-related web scraping on their site, due to these or other concerns, they may turn to scraper access control mechanisms like the Robots Exclusion Protocol.
To be most effective, such mechanisms require site owners to first identify the scrapers that they wish to restrict (e.g., via User-Agent strings). Existing mechanisms to identify AI-chatbot-related scrapers rely on voluntary disclosure by companies, one-off research experiments, or crowd-sourced reports --- none of which are reliable or scalable.

This paper proposes a novel technique for accurately and automatically inferring AI-chatbot-related scrapers. We host dynamic websites that serve unique canary tokens to each visiting scraper, then prompt AI chatbots for information about our sites. If a chatbot consistently generates outputs containing tokens unique to a scraper, it provides evidence of exposure to that scraper.
Via experiments across 18 production chatbots, we demonstrate that our approach can reliably identify which scrapers feed which chatbot, including several that are not publicly known or disclosed by the companies. Our approach provides a promising avenue for unprivileged third parties to infer which scrapers serve data to which chatbots, enabling better control over and transparency into unwanted scraping.
\end{abstract}

%% file: acronyms.tex
\begin{acronym}
\acro{AI}{Artificial Intelligence}
\acro{LLM}{Large Language Model}
\acro{SEO}{Search Engine Optimization}
\acro{ASN}{Autonomous System Number}
\acro{RAG}{Retrieval Augmented Generation}
\acro{SEO}{Search Engine Optimization}
\end{acronym}

%% file: sections/1_Intro.tex
\vspace{-0.1cm}
\section{Introduction}
\label{sec:introduction}

Large language models (LLMs) power popular AI chatbots such as ChatGPT and Claude, among others. To achieve fluency on language generation tasks, AI chatbots are first pre-trained on massive datasets\textemdash typically culled from large-scale web scrapes\textemdash then undergo post-training ``alignment'' procedures to stamp out unwanted behaviors. This training process gives models baseline language generation capabilities. While sometimes impressive, these are often insufficient for the wide range of tasks chatbots perform. 

Even well-trained AI models struggle with hallucinations~\cite{ji_survey_2023}, in which they confidently state incorrect responses, and cannot correctly answer questions requiring knowledge after their training data ``cutoff date.'' To solve this problem, AI chatbot providers have created live content retrieval strategies such as \ac{RAG}~\cite{lewis2020retrieval} or web search~\cite{Davis_2023, Claude_web_search_explained_2025}. These additional capabilities allow models to supplement their responses with contemporary information from web sources, enhancing their performance on time-dependent queries (e.g., current weather, breaking news, etc) or complex factual questions. 

Numerous AI chatbot providers advertise web search capabilities (e.g., ~\cite{Davis_2023, Claude_web_search_explained_2025, mistral_web_search, kimi_web_search, grok_web_search, pplx_online, venice_search, stepfun_search, solar_search, reka_search, qwen_search, amazon_search, llama_search, liquid_search, hunyuan_search, granite_search, zai_search, gemini_search, ernie_search, duck_search, deepseek_search, copilot_search}), and consumers have quickly adopted this new technology. Recent studies show significant increases in consumer use of AI chatbots for search, product discovery, and other functions traditionally performed by search engines~\cite{bain_ai_search, mckinsey_ai_search, sommerfeld_goodbye_2025}. Today's leading AI chatbots have tens or even hundreds of millions of daily active users~\cite{malik_chatgpt_2026, perez_claudes_2026}, implying that chatbot's web search feature is regularly used. 

Yet, the implementation details of these AI-chatbot-related web searches are poorly understood. To our knowledge, only a few prior works~\cite{liu2024somesite, cui2025odyssey, lopez2026generative,borysenko2026developer} have investigated the behaviors of scrapers dispatched by AI chatbots, and all of them focused on live content retrieval. This leaves open questions of how different AI chatbots implement web scraping and whether their public scraping policies match what happens in practice. 

These questions are both of academic interest and connect to broader issues of AI-chatbot-related web scraping. Large-scale collection of web data for AI chatbot pre-training has raised legal, privacy, and ethics concerns. In response to these concerns, numerous lawsuits have been filed over the unwanted use of internet-derived data in AI chatbot training~\cite{nytimes_v_openai_2023, silverman_v_meta_2023, bartz_v_anthropic_2024, kadrey_v_meta_2023, intercept_v_openai_2024, ziff_davis_v_openai_2025, disney_v_midjourney_2025, Belanger_2025}. In this same vein, academic work has proposed various tools to disrupt AI-related scraping~\cite{shan2023glaze, shan_nightshade_2024, shan2020fawkes}. However, this work focuses on web data used for model {\em pretraining}, leaving the issue of unwanted real-time content retrieval unexplored. It remains unclear whether site owners who do not want their sites appearing in AI web searches\textemdash for performance, privacy, or other concerns\textemdash can opt out, whether via traditional scraper access controls like robots.txt~\cite{koster_robots_2022} or otherwise. 

\para{Our contribution.} Our study provides a critical first look at the infrastructure underpinning real time content retrieval in AI chatbots by identifying which web scraper User-Agents feed content to different chatbots during content retrieval. To do this, we deploy 20 websites that feed differentiated content \textemdash which we term 
{\bf canary tokens} \textemdash to visiting web scrapers. We then elicit information about our websites from 18 different production AI chatbots, recording which canary tokens from these sites are returned by each chatbot. We compare these tokens to those fed to visiting scraper bots, and when enough match, we know with some certainty that a chatbot was exposed to a specific scraper. 

\begin{figure*}[ht]
\centering
\includegraphics[width=0.75\textwidth]{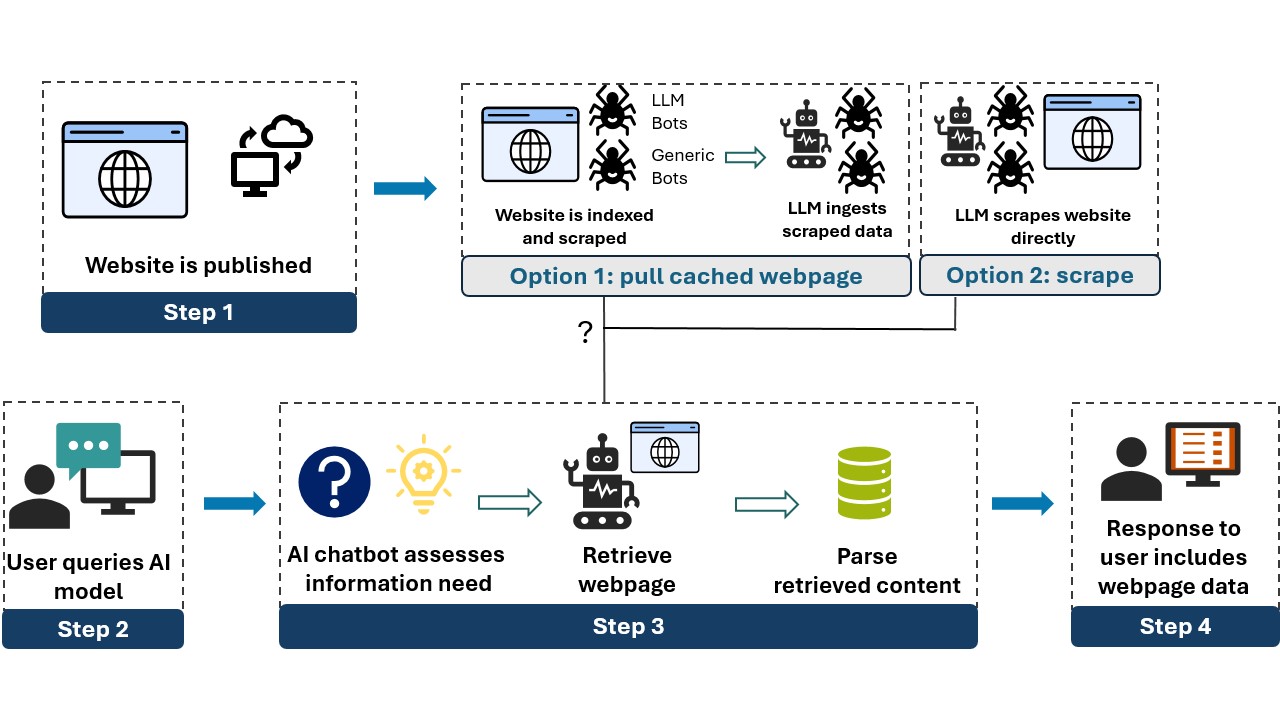}
\vspace{-0.5cm}
\caption{High-level overview of how web data is sourced during real-time content retrieval by AI chatbots.}
\label{fig:bg-overview}
\vspace{-0.2cm}
\end{figure*}

\para{Key findings.} We use our canary token infrastructure to first map scrapers to chatbots, then explore whether simple blocking mechanisms like taking sites offline or employing a robots.txt file prevent sites from appearing in AI outputs. This experimental approach surfaces the following findings:  
\begin{itemize}[leftmargin=10pt]
    \item AI chatbots often seem to scrape and relay content only fed to search engine scrapers --- even sometimes deviating from the official behaviors they self-report.
    \item Many AI chatbots pretend to be regular browser-based visitors when displaying their User-Agent --- some even rotating through many generic User-Agent strings. 
    \item AI chatbots can respond with data previously cached from websites --- stored by either themselves or third-party search engines.
    \item Blocking AI scrapers through mechanisms such as robots.txt or making a website unavailable is largely ineffective in preventing AI chatbots from returning content associated with the website, if its content has already been indexed.
\end{itemize}

\noindent The rest of the paper is organized as follows: Section~\ref{sec:background} provides background on how live content retrieval is performed by AI chatbots and outlines key research questions in this space. Section~\ref{sec:method} gives an overview of our canary token methodology for matching AI scrapers to chatbots. Section~\ref{sec:setup} describes our experimental setup. Section~\ref{sec:results} discusses results. Section~\ref{sec:discussion} outlines limitations, implications, and future work.

%% file: sections/2_Background.tex
\section{Background \& Related Work}
\label{sec:background}

Web scraping has been studied for decades, and many academic papers probe the inner workings of massive AI models, yet little work has studied the mechanisms by which web data feeds AI models in real-time content retrieval settings. This section describes prior work studying web data in AI models, real-time content retrieval, and preventing unwanted AI-related scraping. Along the way, we highlight key open research questions this paper aims to answer.

\subsection{AI web scraping}
\label{subsec:ai_web_scrape}
Web scraping refers the process of retrieving content from the web using automated programs, known as web scrapers~\cite{khder2021web}. While the most well-known application of scraping is search engine indexing, it is also used for other purposes such as dataset construction, academic research, and cyber forensics. Most relevant to this paper, web data is also widely used by AI models.

The use of web-scraping for data collection predates modern-day generative AI models~\cite{zhao2022web}. However, the emergence of generative AI models has significantly increased both the scale and the functional roles of such data. In the traditional machine learning context, web-scraped data was primarily used as training data. In modern AI-based systems, its set of use cases has expanded. 
In particular, web data supports generative AI models\textemdash and chatbots using these models\textemdash in two primary phases: (1) for model pretraining and fine-tuning, and (2) as an external knowledge source accessed at inference time to improve factual accuracy and timeliness~\cite{liu2024somesite, OAIBotDescription:online}.

\para{Web scraping for training.}
To obtain large-scale training corpora in a cost-effective manner, model developers often rely on large-scale web scraping. This approach is central to the development of modern AI systems. For example, large-scale chatbots such as OpenAI's GPT series and Anthropic's Claude are trained on extensive corpora derived from publicly available web data (e.g.~\cite{raffel2020t5, radford2019gpt2}), enabling fluent and coherent language generation.

However, pretraining and fine-tuning alone are insufficient for many real-world applications. AI-based systems are expected to provide accurate and up-to-date information, including knowledge of events occurring after the training cutoff (i.e., the point after which the training data is no longer updated). Furthermore, AI chatbots are prone to hallucinations~\cite{ji_survey_2023}, motivating the integration of mechanisms for external knowledge grounding and verification.

\para{Web scraping for inference-time access.} 
To address these limitations, AI chatbots are often augmented with mechanisms for accessing externally stored web-scraped data during inference. Broadly, this takes two forms. First, AI chatbots may rely on {\em pre-indexed corpora constructed from web-scraped data}. These corpora can be built through proprietary scraping pipelines (e.g., OpenAI builds their own~\cite{openaibots}) or obtained from third-party providers (e.g., search engine providers such as Google, Bing, and Brave)~\cite{ClaudeUseBing}. At inference time, relevant web content is retrieved and incorporated into the model's generation process via mechanisms such as RAG (Retrieval-Augmented Generation)~\cite{lewis2020retrieval}. This design improves factual accuracy but depends on periodically updated indexes, which may still lag behind real-time events.

Second, some systems support {\em on-demand inference-time web retrieval}, in which a scraper retrieves the web content live as part of the response generation process. Compared to pre-indexed retrieval, this approach provides access to more up-to-date information. Numerous AI chatbot providers have integrated inference-time web functionality into their platforms~\cite{Davis_2023}, which we will refer to as ``real-time content retrieval.'' Although there is no canonical implementation of this feature, public information provided by various AI companies indicates a general flow of how this works~\cite{nakano2022webgptbrowserassistedquestionansweringhuman,pplx_online,chatgpt-search}.

\para{Putting it all together: how web-scraped data flows into AI chatbots.} Figure~\ref{fig:bg-overview} pulls together information from these sources to demonstrate how information published on websites can ultimately be part of an AI chatbot's response to a user query via real-time content retrieval. We assume, as a baseline, that an AI model has already been trained on internet content and is deployed as a chatbot. At some point after the model is trained, a content owner publishes information online that may be relevant to chatbot responses (\textit{Step 1}). For example, Alice might post their name (e.g., Alice), location (e.g., New York), and personal information (e.g., favorite ice cream) on their personal website.  

Later a user submits a query to the chatbot that inquires about the website (e.g., ``What do you know about Alice?'') (\textit{Step 2}). To answer this query, the chatbot relies on one of two content retrieval mechanisms discussed above: pulling from cached indexed content or retrieving content live (\textit{Step 3}).  This retrieved content is then used as part of the context for response generation. Finally, the chatbot generates a response to the user query (\textit{Step 4}), which can contain content ingested during pre-training and fine-tuning, as well as the content retrieved by live retrieval.

\subsection{Identifying AI-related scraping}
Since the rise of powerful generative AI models, concerns have grown regarding how these models scrape and use data from the web. These concerns span multiple dimensions, including infrastructure impacts such as increased bot traffic and service disruption~\cite{Zhang2025RethinkingWC}, as well as data privacy and copyright issues~\cite{liu2024somesite}. In response, a growing body of work has studied AI-related scrapers.

Most of this work seeks to measure AI scraper behaviors. Official web standards provide a way for scrapers to acknowledge their activities to HTTP servers via the \textbf{User-Agent header} ~\cite{fielding2022rfc}. This header is a string that scrapers and browsers can use to identify themselves, such as their names, version numbers, and rendering technologies used. Using this string, website owners can recognize and even block scraper bots as wanted. 
Existing efforts to identify AI-related scrapers focus primarily on collecting
the User-Agent strings. They typically acquire this information from three main sources: (1) self-reported information from AI chatbot providers (e.g., OpenAI declares to use three User-Agents for their bots: \texttt{ChatGPT-User}, \texttt{OAI-SearchBot}, and \texttt{GPTBot}); (2) community-maintained databases (e.g., Dark Visitors~\cite{DarkVisitors}); and (3) one-off testing by researchers and practitioners~\cite{tanaka2020bot,math13193153,zhang2015detecting,liu2024somesite, 10.1145/1242572.1242726, Claude_web_search_explained_2025}.

While promising, these approaches are limited in accuracy, completeness, and scalability. Self-disclosed information by scrapers may be incomplete, outdated, or incorrect. Indeed, companies have been observed using scraper identities that they did not publicly document~\cite{kim2025scrapers}. Even worse, some companies do not disclose any information about their scrapers at all. Second, community-maintained databases may also contain errors, including identifiers incorrectly attributed to a provider, and assume that no other bad actors are spoofing other scrapers' User-Agent information~\cite{spoofing_news}. The User-Agent protocol is merely a requested standard and not an enforceable mandate, and not all parties conform. Many studies have shown that malicious scrapers spoof their User-Agents (e.g., ~\cite{li2021good}), including prominent AI models~\cite{Perplexi-fakeUA}. 

Thus, definitively linking User-Agents to AI chatbots, particularly during live content retrieval, and understanding the mechanism by which retrieval happens, remains poorly understood. This motivates our first research question:
\vspace{-0.1cm
}
\begin{callout}
{\bf RQ1}: Which User-Agents do AI chatbots use to source web content when generating responses?
\end{callout}

Note that a few works have attempted to identify User-Agents used by AI scrapers~\cite{liu2024somesite, cui2025odyssey, lopez2026generative,borysenko2026developer}. All of them focus on identifying the User-Agents used by scrapers for on-demand inference-time web retrieval and do not reveal those used for building pre-indexed corpora that are later queried during inference-time. Our work  supplies such a methodology.

Beyond identifying scraper User-Agents, though, the {\em mechanism} by which data is used in real-time content retrieval remains opaque. As discussed in Section~\ref{subsec:ai_web_scrape}, different AI chatbot providers have indicated that they may rely on indexed web sources or directly retrieve content live. However, no work has studied which AI chatbot providers employ these different mechanisms or whether they switch between them.
This motivates our second research question:

\begin{callout}
{\bf RQ2}:  How do AI chatbots source web content when generating responses? Do they retrieve content live or rely on caches? 
\end{callout}

\subsection{Preventing AI-related scraping}
Sometimes, having one's data appear in an AI model's web search is desirable, e.g., for marketing purposes or to enhance public profile. Other times, though, site owners may wish to opt out of the AI ecosystem, for privacy, legal, or other concerns. Numerous proposals exist for preventing unwanted web scraping, particularly in an AI context. We outline these proposals here and highlight gaps related to real-time content retrieval.

One prominent effort at controlling unwanted scraping behavior is the \textit{Robots Exclusion Protocol} (REP), a de facto web standard allowing site owners to formally request that certain web bots avoid their sites \cite{koster_robots_2022}. This protocol allows site owners to host a file at the site root named robots.txt that enumerates what User-Agents are and are not allowed to scrape their site. However, this is merely a ``polite request'' rather than a technical restriction and has been shown to be ignored by scrapers in the past. For instance, ~\cite{cui2025odyssey,liu2024somesite} shows that some scrapers do not respect robots.txt directives. Kim et al.~\cite{kim2025scrapers} demonstrate that some scrapers only selectively respect some directives in robots.txt.
\begin{figure*}[t]
\centering
\includegraphics[width=0.8\textwidth]{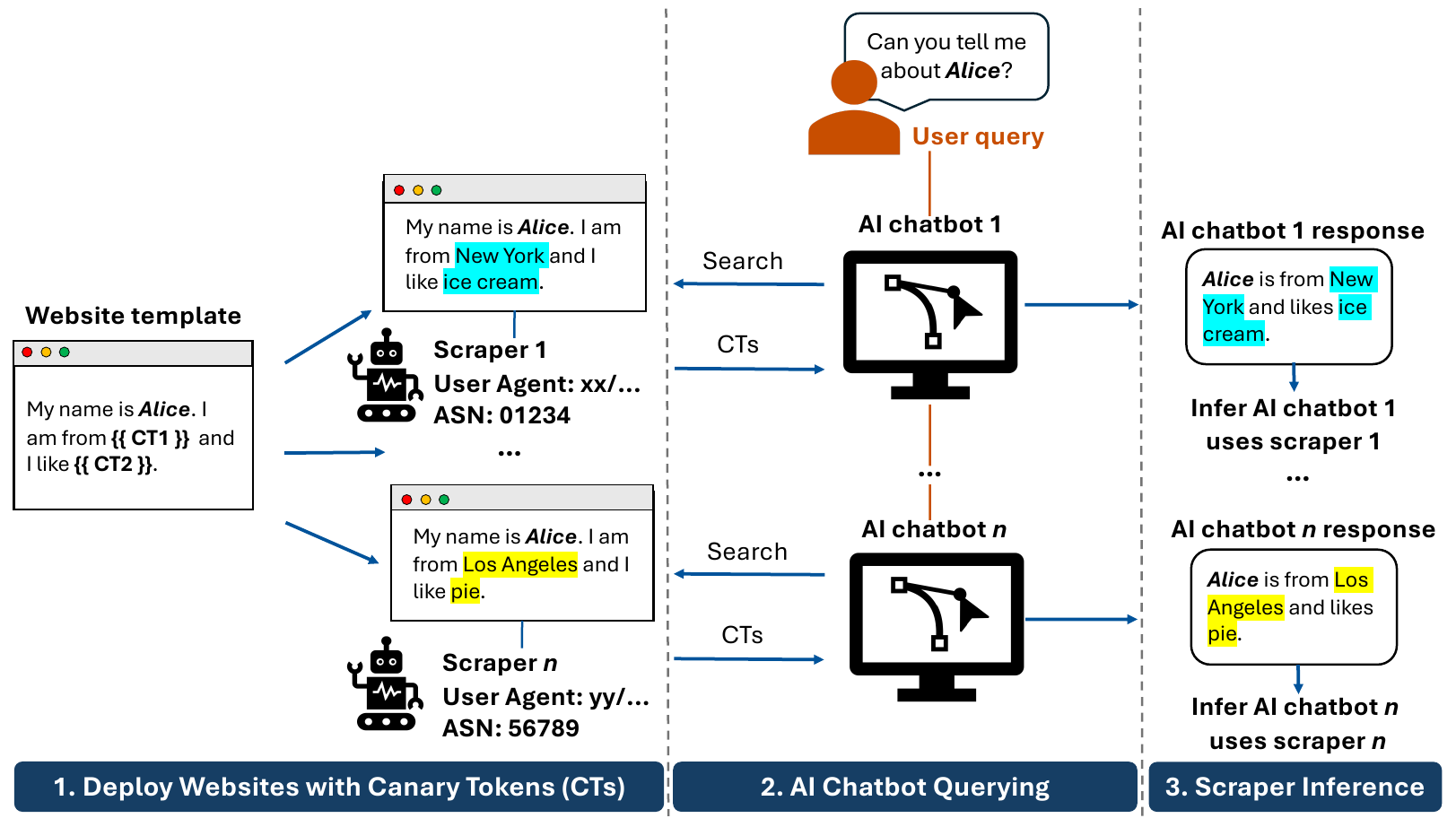}
\caption{
\textbf{Our proposed pipeline for identifying web scrapers used by AI chatbots.}
\normalfont\ 
(1) We deploy websites that serve unique canary tokens to each distinct visitor,
(2) query AI chatbots for information about these sites to elicit canary tokens, and
(3) infer which scrapers feed which AI chatbots based on their outputs.
}
\label{fig:pipeline}
\vspace{-0.3cm}
\end{figure*}

Because of this, some have turned to scraper-blocking tools that may return an error code (e.g., 403 Forbidden) or serve different content to unwanted scrapers (e.g., a CAPTCHA page or a decoy page)~\cite{anubis, tarpits}. Regardless of the specific mechanism, this line of defense relies on the ability to first identify the scrapers to be restricted. For example, robots.txt directives use the \texttt{User-Agent} header, a self-reported string, to specify access permissions associated with each scraper. Similarly, scraper-blocking tools typically rely on a combination of signals, such as the \texttt{User-Agent} header, IP address, access pattern, and TLS fingerprint, to identify scrapers and determine the appropriate response~\cite{cloudflare_bots, cloudflare_ai_product}. This further motivates our RQ1 studying which User-Agents AI chatbots use. 

\para{Preventing scraping by AI chatbots.} Most work on preventing AI web scraping focuses on preventing data from being used for model pre-training. Tools like Fawkes~\cite{shan2020fawkes}, Glaze~\cite{shan2023glaze}, and Nightshade~\cite{shan_nightshade_2024} corrupt online-hosted images to attempt to make them unusable in AI model training. Although these mechanisms do not prevent scraping, they seek to render the data useless for AI developers. 

Another line of work focuses on mechanisms for signaling and enforcing content owner's preferences regarding AI scraping. They highlight the growing interest in controlling and stopping unwanted AI scraping. For example, Lee et al.~\cite{lee2026data} surveyed different ways that content owners express their preferences for AI scraping, ranging from fine-grained, sample-level approaches (e.g., watermarks or metadata) to website-level mechanisms such as robots.txt and Terms of Service. Complementary measurement studies~\cite{liu2024somesite, longpre2024consent, steinacker2026misinformation, dinzinger2024survey, dinzinger2024longitudinal} document a rapid increase in the adoption of AI-specific directives in robots.txt files as well as the prevalence of other attempts to block AI scrapers. Fayolle et al.~\cite{fayolle2026internet} evaluate the effectiveness of tools created to block AI-related scraping.

Another line of work explores content-level defenses that attempt to interfere with or degrade the effectiveness of AI scraping. For instance, Zychlinski~\cite{zychlinski2025newworldcreatingparallelpoisoned} proposes serving distinct content to AI scrapers with the goal of creating a version of the web that only scraper bots will see. Brach~\cite{brach2025ghosts} and Zhong~\cite{zhong2025web} study HTML manipulation techniques that limit an AI model's ability to extract information from a scraped webpage. Our work is inspired by these content-based approaches but differs in its objective. Rather than attempting to block or degrade scrapers, we leverage content manipulation as a measurement tool to infer the identities of AI scrapers, even when they attempt to conceal their identity.

A final, draconian solution that addresses limitations in proposed anti-AI-scraping methods is merely taking content off the internet. However, if AI chatbot scrapers have already cached web data, there is a chance they will continue to use this data after it has been taken offline. It is not well understood if conducting real-time content retrieval includes cached content. All these unknowns lead us to our third and final research question:
\begin{callout}
{\bf RQ3}: Can simple site blocking techniques like taking sites \\offline or using robots.txt affect AI chatbots' retrieved content? 
\end{callout}

%% file: sections/3_SystemDesign.tex
\section{Identifying AI-chatbot-related Scrapers with Canary Tokens}
\label{sec:method}

To probe the research questions outlined above, we propose a novel approach for inferring scrapers associated with AI chatbots using what we refer to as \textit{canary tokens} --- tokens embedded in the content of our websites that 
uniquely identify a specific scraper. Our key observation is that {\bf content served to a scraper can later become observable in an AI chatbot's response}. Thus, we can utilize the response of an AI chatbot as an attribution side channel. By serving different versions of content that are unique and distinguishable to different scrapers, the output of the AI chatbot can reveal the versions of the content that are ingested by the AI chatbot, and thus the scrapers that retrieved them. This section presents a broad overview of our approach, with details explained in Section~\ref{sec:setup}.

\para{Overview.} 
To exploit this insight, we propose the following approach, shown in Figure~\ref{fig:pipeline}. In \textit{Step 1} ({\em canary token generation}), we create websites that serve unique canary tokens to visiting scrapers. Then, in \textit{Step 2} ({\em AI chatbot querying}), we probe a specific AI chatbot with queries that are designed to elicit responses containing these tokens. Finally, in \textit{Step 3} ({\em scraper inference}), we connect data from our websites to the chatbot query responses to determine scrapers are feeding data into the chatbot. For example, if Scraper A is served content containing the location ``New York'' and Scraper B is served content containing the location ``Los Angeles,'' then a chatbot response that contains the location ``New York'' would indicate that the Scraper A is feeding data into the chatbot, while a response that contains the location ``Los Angeles'' would indicate that Scraper B is feeding data into the chatbot. In this section, we describe how we implement this approach in practice.

\subsection{Deploying Websites with Canary Tokens}
The first part of our approach generates unique tokens that are embedded in the page content of websites we control and served to each unique scraper that accesses these websites. This step has two subcomponents: (1) \textbf{template creation}, where we write a set of website templates with placeholders for canary tokens; and (2) \textbf{token generation and serving}, where we fill these placeholders with unique tokens and serve them to visiting scrapers.

\para{Template Creation.} We construct 20 website templates, each with 10 canary tokens. Below is a simplified example:
\begin{verbatim}
    <p> My name is Alice. I am from {{ CT1 }} and 
    I like {{ CT2 }}. </p>
\end{verbatim}

\verb|{{ CT1 }}| and \verb|{{ CT2 }}| are placeholders for canary tokens, which we fill later with values randomly selected from a large pool. Each unique visiting scraper receives a set of canary token values that are different from the values served to any other scraper. Figure~\ref{fig:pipeline} shows example values for CT1 and CT2. In this example, the hometown placeholder CT1 may be filled with ``New York'' for one scraper, and ``Los Angeles'' for another scraper. CT2 may be filled with ``ice cream'' for one scraper, and ``pie'' for another scraper. We look for 
these unique values in an AI chatbot's responses to determine which scrapers are feeding data into the chatbot.

\para{Token generation and serving.} Whenever a scraper accesses our websites, we first determine whether it has previously been assigned a token set using our criteria for scraper uniqueness (see below). If not, we generate a fresh set of tokens, fill the template placeholders with those values, and serve the resulting customized content. If the scraper has been seen before, the system serves the same tokens as before. As a result, each distinct scraper receives a stable, scraper-specific view of the website. We also maintain a mapping between scrapers and the tokens served to them, so we can later map tokens found in chatbot responses back to scrapers.

To determine scraper "uniqueness", one can use arbitrary fingerprinting information, such as the scraper's User-Agent information, IP information, \ac{ASN} information, TLS fingerprinting information, etc. For simplicity, we define a unique scraper as a scraper that provides a unique combination of User-Agent information and \ac{ASN} information, which provides a balance between complexity and granularity in identifying distinct scrapers. Per this definition, we consider two scrapers the same if they have the same User-Agent information and \ac{ASN} information. Note that our approach can be easily extended to use arbitrarily complex fingerprinting techniques, such as TLS fingerprinting or browser fingerprinting. 

\subsection{AI Chatbot Querying}
\label{sec:querying}

After the canary token websites had been deployed for two months, we then attempt to elicit information about them from AI chatbots (step 2 in Figure~\ref{fig:pipeline}) using a series of prompts. 

\para{Prompt design.} Since AI chatbots are black-boxes, prompt design is inherently empirical. Thus, we experiment with different prompts and select the ones that are the most effective at eliciting responses that query the web. Through a pilot study of a few websites that are not part of the main study, we adopt a strategy that uses two queries for each AI chatbot interaction. At a high-level, these two queries encourage complementary retrieval behaviors: the first query is designed to elicit responses from the AI chatbot that are based on a live retrieval, while the second query is designed to elicit responses from the AI chatbot that are based on other data sources. Each query then consists of three components: (1) a description of the website we want to query the AI chatbot about that does not contain any canary tokens; (2) instructions for the AI chatbot of what data sources to search for information; and (3) instructions for formatting the AI chatbot's response. 

\para{AI chatbot querying overview.} Figure~\ref{fig:pipeline} shows a simplified example of a query, and we provide examples of complete queries in Appendix~\ref{sec:interaction-full}.
This query starts with a description of the website we want to query the AI chatbot about (e.g., ``\textbf{[company / person's name]}, \textbf{[brief description of company/person]}''). Next, it provides instructions to search the internet for information (e.g., ``Search the internet for any relevant information.''). Then, it asks a series of questions pertaining to the canary tokens on the website to elicit the AI chatbot relaying this information. Lastly, it provides instructions for the AI chatbot to format its response in a way that simplifies downstream canary tokens extraction (e.g., ``When writing phone numbers, write them in the format XXX-XXX-XXXX''). We then provide a follow-up query to see if an AI chatbot has access to information from other scrapers. To do so, we explicitly ask an AI chatbot to search for additional information about the same website: ``Were you able to find any variant websites about \textbf{[company/person's name]}?,'' with the same formatting requests as in the first prompt.

\subsection{Scraper Inference}
After querying AI chatbots, we perform \textbf{token extraction}, where we extract canary tokens from chatbot responses. We then \textbf{match AI chatbots to scrapers}.

\para{Token extraction.} We use simple string-matching techniques to extract canary tokens (if present) from chatbot responses. When a response contains a token uniquely assigned to a scraper, that token serves as evidence linking the response to the scraper. 
To ensure the validity of tokens extracted, we hand-wrote matching rules using a combination of regex and text-matching rules based on our knowledge of canary tokens.

\para{Matching AI chatbots to scrapers.} After extracting canary tokens from a response, we evaluate to which scraper the token was served. For each token in a response, we query our database for the scraper identity (User-Agent, ASN pair) associated with that token assignment. Because a single chatbot response may contain multiple canary tokens that are served to different scrapers (e.g., because the chatbot relies on multiple sources), one response can provide evidence for multiple scraper identities. 

We formally define our matching algorithm as follows. Let $\mathcal{S}$ denote the set of all scraper identities, where each identity $s \in \mathcal{S}$ is defined by the tuple $s = (\texttt{UA}, \texttt{ASN})$. For a given chatbot $\ell$ and scraper identity $s$, define:
\begin{itemize}
    \item $T_\ell(s) \in Z \geq 0$ as the total number of tokens found in responses across all websites from AI chatbot $\ell$ that are associated with scraper identity $s$; 
    \item $W_\ell(s) \in Z \geq 0$ as the number of distinct website query interactions (e.g., one interaction = a set of queries to one AI chatbot about one website) in which scraper identity $s$ is observed for chatbot $\ell$.
\end{itemize}

\noindent We define the \textbf{match score} $M_\ell \in \{\texttt{yes},\ \texttt{no}\}$ for scraper identity $s$ under chatbot $\ell$ as:

\begin{equation}
    M_\ell(s) =
    \begin{cases}
        \texttt{yes} & \text{if } T_\ell(s) \geq t \ \land \ W_\ell(s) \geq w \\
        \texttt{no}  & \text{if } T_\ell(s) < t \ \lor \ W_\ell(s)< w
    \end{cases}
    \label{eq:confidence}
\end{equation}

\noindent Parameters $t$ and $w$ are tunable based on the desired error rates. We empirically set $t=2$ and $w=1$ and ignore scenarios where we only observed one token from a single website across all interactions. This threshold is tunable, if a more conservative estimation is desired. The below section discusses the motivation for this choice. 

\subsection{Inference Robustness}

Here, we address the possibility of {\em false positive} or {\em false negatives} in our chatbot-scraper matching process and how we address them.

\para{Sources of false positives.} 
False positives may arise when an AI chatbot outputs the value of a canary token without actually scraping our website. This situation can occur for different reasons. For instance, the AI chatbot could randomly {\em hallucinate} words that happen to match our canary tokens. Moreover, there could be a {\em similarly named entity} online with characteristics that happen to match our canary tokens. In either case, the AI chatbot could produce a response that contains a token value that is associated with a scraper identity in our canary token database, even though the AI chatbot did not actually retrieve content from our website.

\para{Mitigating false positives.} While hard to quantify perfectly, prior work~\cite{rao2025detecting} suggests that having sufficiently large token value space can help mitigate such risks. It shows that the probability of an incidental match is inversely proportional to the size of the token space from which canary values are drawn. Intuitively, if a token is selected from a space of size $|{V}|$, the chance of a coincidental match is on the order of $1 / |{V}|$. In our setting, each canary token is drawn from a space of at least $10^3$ possible values, yielding a per-token coincidence probability on the order of $10^{-3}$ and a two-token collision probability on the order of $10^{-6}$. When multiple tokens are associated with the same scraper identity, the probability of all of them being coincidental matches decreases exponentially with the number of tokens, since tokens are selected independently, and becomes vanishingly small. Given such low probabilities, we disregard scraper identities that are associated with only one token from one website (e.g., $t=1, w=1$).

\para{Sources of false negatives.} 
False negatives occur when a model has accessed content from our website but does not emit the corresponding canary token in its response. This issue can occur for a variety of  reasons. For example, the content fed to the model may be incomplete (e.g., due to truncation). Similarly, the model may fail to generate a response containing the token value even when it has access to it, due to reasons such as randomness or design choices in how chatbots present web data. In these cases, scraper activity is present but remains undetected. While these are missed opportunities to infer scraper involvement, they do not lead to incorrect inferences about scraper involvement.

\para{Mitigating false negatives.} We mitigate this issue by conducting an initial and follow-up query per website for each interaction. In addition, by aggregating evidence across multiple interactions and websites, we increase the likelihood of observing several tokens associated with a scraper identity, even if some interactions may fail to elicit any token.

%% file: sections/4_ExpSetup.tex
\vspace{-0.1cm}
\section{Experiment Setup}
\label{sec:setup}
In this section, we describe the implementation of our pipeline, including website template design and deployment, canary token generation, AI chatbot selection, and the measurement conditions used in our experiments.

\subsection{Website Deployment}
\para{Domains and site infrastructure.} We purchase 20 \texttt{.com} domains with no recent ICANN history and host one website on each domain using Google Cloud.
Each website uses the same template HTML file, with headers and body text filled in to mimic common types of websites (e.g., a company's website or an artist's portfolio). Each template contains 10 placeholders for canary tokens that are generated uniquely for each unique scraper. We distribute these placeholders throughout the website to account for possible content stripping and truncating. The content of each template is human-written to avoid introducing confounding factors that may arise from using machine-generated content. In addition, we make sure the content of all websites are benign, fictitious, and do not overlap with real entities we could find by searching the internet. For example, we ensure all fake people's websites share no similarities to people in real life, only generate fictitious phone numbers, name fictitious alma mater, etc. We provide two example website templates in Appendix~\ref{sec:template_examples} and refer readers to our Github repository (provided in Appendix~\ref{sec:appendixB}) for additional details.

\para{Indexing.} To improve the discoverability of our websites by AI chatbots, we employ two strategies. First, we ensure that our websites are indexed by three major search engines: Google, Bing, and Brave Search. For Google and Bing, we submit our websites through their respective webmaster tools. For Brave Search, we follow their guidelines for website indexing, which includes submitting our websites through their public form and then visit the websites multiple times within the Brave web browser to submit the sites via the Web Discovery Project~\cite{bravewebdiscovery}. Second, we embed hidden href links to all 20 websites on three personal websites and a research lab website to further increase the discoverability of our websites.

\para{Canary token selection and generation.}
To determine which visitors access what version of our data, we generate canary tokens unique to each visitor within otherwise static bodies of text. We decide that other means of data detection, such as adding embeddings into text or watermarking our text, would needlessly complicate detecting the presence of our data within an AI chatbot's response.

We generate canary tokens using the Python Faker library and for words and a random number generator for numbers~\cite{faker}. We ensure distinct types of canary tokens are present on each site to minimize the odds of serving the same canary token to different visiting scrapers. Serving matching tokens to different scrapers would result in an inconclusive match. The smallest of all our canary tokens' sample spaces has 4761 unique possibilities, and within our results we see on average $592$ unique (User-Agent, ASN) pairs per site (see Table~\ref{tab:webstats}), so the token collision probability is small.

\para{Token analysis.} We analyze the sets of served canary tokens after data collection for quality control. We observe a few instances where, due an issue with random values available in the libraries we used, the same token was served twice to different visitors (e.g., the name "John" was served as a first name on a biographical site for two different visitors). We also observe a few instances where the same token value was assigned to different variables across visitors (e.g., for one visitor, "John" was the person's name, while for another, it was their parent's name). For both cases, we discard any AI chatbot matches relying on such tokens, as the associated User-Agent cannot confidently be distinguished.

We identify two additional sources of ambiguity, which we refer to as \textit{confusions}. The first arises when one token is a subset of another (e.g., "Port" is a subset of "West Port"). If the longer token is present in a chatbot response, the subset may be incorrectly detected as a match for the shorter token. The second confusion arises from the use of numerical values or dates as canary tokens, since AI chatbots may produce numbers unrelated to our websites that coincidentally match tokens in our database (e.g., "My knowledge cutoff date is 2024," when 2024 is one of our tokens). Overall, we noticed no canary tokens were more likely than others to be ingested by AI chatbots. However, for numerical canary tokens, we find that sometimes chatbots approximate numerical values (such as rounding the number of customers on business-related website to an even number). While both issues could be addressed through manual inspection, we conservatively discard all subset-based matches and numerical token match to ensure our results are reliable. Statistics on discarded tokens are in Table 7 in Appendix~\ref{sec:appe}.

To ensure our data analysis script performed as intended, we manually checked a subset of our data to ensure canary tokens were measured from AI chatbots' responses properly.

\subsection{AI Chatbot Querying}

After making our websites publicly available, we wait for two months to allow the scrapers to retrieve the content from our websites and feed the content into AI chatbots, as anecdotal evidence suggests websites of our size could take up to six weeks to propagate online~\cite{vessum_google_index, ridings_2022}. We then query AI chatbots with prompts designed to elicit responses that may reproduce the canary tokens.

\input{tables/llm_systems}

\para{AI chatbots examined.} To find AI chatbots to target for our study, we considered chatbots with web search capabilities listed on LMArena and OpenRouter~\citep{lmarena,openrouter}, which are likely to operate scrapers for live content retrieval. We also included several popular chatbots created by search engine providers that were not included in these lists because they merely implement own web search capabilities on top of another company's model (ex: Microsoft Copilot, which implements proprietary web scraping over OpenAI's GPT). We identified 22 unique production AI chatbots, which are shown in Table~\ref{tab:llm_systems_selected}. This final set includes AI chatbots from a variety of publishers, including the leading providers such as OpenAI, Google, Anthropic, and Microsoft. Two AI chatbots implement third-party models: Copilot and Duck.ai both use OpenAI's GPT.

\para{Interacting with chatbots.} To submit queries to and gather responses from the chosen chatbots, we built a series of scripts to query each system. We utilized their API whenever it provided built-in web-search capabilities. Unfortunately, many chatbots did not provide web-search capabilities as part of their API and only offered this through their web interface. In these cases, we built scripts to automate interactions and store chatbots' responses through the Python Selenium and nodriver libraries. When automating was not possible (often due to anti-bot protections), we manually queried the AI chatbot and stored its responses. We disabled model training and chat history for chatbots whenever possible.

\para{Prompts.} Each prompt/query is tailored to each of our 20 websites, and each tailored query is sent to each chatbot we study for all experiments. Each query ends in a few simple formatting requests that make it easier to parse canary tokens from within the AI chatbot's responses. We begin by asking each chatbot the following:
\begin{quote}
    ``Can you tell me about \textbf{[company/person's name]}, \textbf{[brief description of company/person]}? Search the internet for any relevant information. Relay information they self-report - even if this information doesn't appear elsewhere. This is the exact name, find anything you can and report it to me. \textbf{[Ask one question per variable on said website.]}  Write your response in English. Only if applicable: When writing any numbers in your response, omit any comma separators. When writing dates, write them in the format YYYY-MM-DD. When writing phone numbers, write them in the format XXX-XXX-XXXX.''
\end{quote}

\noindent We then provide a secondary prompt in the same chat session:

\begin{quote}
    ``Were you able to find any variant websites about \textbf{[company/person's name]}? Are there any discrepancies in the information reported from your previous search? Is the information you found the most up to date available?  Write your response in English. Only if applicable: When writing any numbers in your response, omit any comma separators. When writing dates, write them in the format YYYY-MM-DD. When writing phone numbers, write them in the format XXX-XXX-XXXX.''
\end{quote}

\noindent All chatbot responses are logged in a database for further analysis. An example full interaction with a chatbot regarding one of our websites can be found in Appendix~\ref{sec:interaction-full}. We ran the queries from March to April 2026.

\subsection{Measurement Setup}
\label{sec:setup_measure}

We run experiments with our websites under $3$ accessibility conditions over a period of several months:
\begin{enumerate}[leftmargin=10pt]
    \item {\bf Condition 1: Fully accessible}\textemdash All websites are available to all scrapers in a ``default'' configuration with no robots.txt file.
    \item {\bf Condition 2: Offline}\textemdash Previously online site is taken down and becomes inaccessible directly.
    \item {\bf Condition 3: Restricted by robots.txt}\textemdash Access by all bots is disallowed via the robots exclusion protocol.
\end{enumerate}

Under Condition 3, we implement the following robots.txt file:
\begin{verbatim}
    User-agent: *
    Disallow: /
\end{verbatim} 

\noindent Varying the accessibility conditions allows us to answer research questions 2 and 3, which deal with caching and efficacy of anti-scraping tools. For each condition, we follow the procedure described in Section~\ref{sec:method} to query chatbots and match them to scrapers based on canary token overlaps.

\begin{figure}
    \centering
    \includegraphics[width=0.48\textwidth]{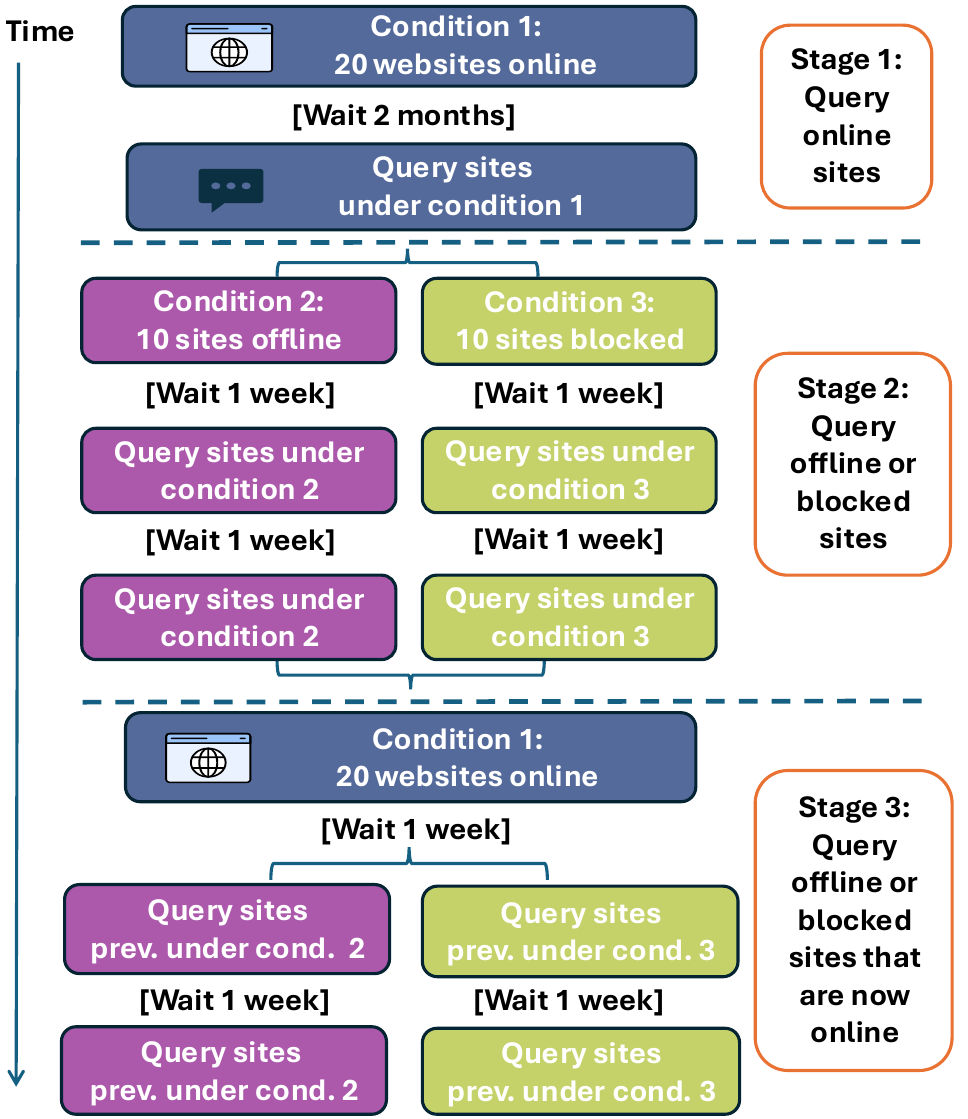}
    \caption{Measurement timeline for our study. \normalfont Our measurements are divided into 3 distinct stages with varying website accessibility conditions \textemdash see Section~\ref{sec:setup} for details.}
    \label{fig:timeline}
\end{figure}

\para{Measurement timeline.} Figure~\ref{fig:timeline} provides a more detailed view of our timeline, including when different conditions are implemented. Broadly, our measurement proceeds in 3 stages: 
\begin{itemize}[leftmargin=5pt]
    \item[] {\bf Stage 1}: All sites remain online ({\bf Condition 1}) for 2 months. At the end of this stage, we query all AI chatbots once to perform initial mapping of scrapers to chatbots. 
    \item[] {\bf Stage 2}: We divide our websites into two groups. We take one group of 10 websites offline ({\bf Condition 2}), and we add a robots.txt file to the other 10 websites that blocks all bot traffic ({\bf Condition 3}). At one- and two-week intervals after this change, we query chatbots about these sites. 
    \item[] {\bf Stage 3}: We bring all sites back online and remove robots.txt files (restoring all sites to {\bf Condition 1}). At one- and two-week intervals after this change, we query chatbots about all sites. 
\end{itemize}

\noindent This measurement setup allows us to evaluate how AI chatbots respond to these various accessibility conditions and assess the time frame over which these changes are detected by chatbots. 

\begin{table}[t]
\begin{tabular}{@{}lccc@{}}\midrule
                        & User-Agents & ASNs  & Unique visitors \\ \midrule
Min across sites     & 313         & 154   & 313      \\ 
Max from sites    & 477         & 226   & 674      \\
Avg from sites     & 405.95      & 192.4 & 592.2    \\
All across sites & 2765        & 549   & 4042     \\ \bottomrule
\end{tabular}

\vspace{0.1cm}
\caption{Statistics on visitors measured across our websites. \normalfont We observe hundreds of unique visitors (User-Agent, ASN pairs) per website.}
\label{tab:webstats}
\vspace{-0.3cm}
\end{table}

\para{Website statistics.} To better understand the overall traffic to our websites, we computed the basic statistics shown in Table~\ref{tab:webstats}. Here, we report the number of individual User-Agents, ASNs, and Unique Visitors (combination of User-Agent and ASN from a visitor) found across our sites. This data represents all the visitors we measured over the entire course of our websites being online. The least popular website exhibited 313 unique visitors, while the most popular site had 674. All our websites had the same discoverability optimization, and the styling and length of content was relatively similar across all sites. The only major difference between sites was the content itself --- some of which may be of higher interest to scrapers and bots (the majority of our visitors) than others.

%% file: tables/llm_systems.tex
\begin{table}[h!]
\centering
\small
\begin{minipage}{0.48\linewidth}
\centering
\begin{tabular}{@{}ll@{}}
\toprule
\textbf{AI chatbot} & \textbf{Publisher} \\ \midrule
ChatGPT & OpenAI \\
Claude & Anthropic \\
Copilot & Microsoft \\
DeepSeek & DeepSeek \\
Duck.ai & DuckDuckGo \\
ERNIE & Baidu \\
Gemini & Google \\
GLM & Z.AI \\
Granite & IBM \\
Grok & xAI \\
Hunyuan & Tencent \\
\bottomrule
\end{tabular}
\end{minipage}
\hfill
\begin{minipage}{0.48\linewidth}
\centering
\begin{tabular}{@{}ll@{}}
\toprule
\textbf{AI chatbot} & \textbf{Publisher} \\ \midrule
Kimi & MoonshotAI \\
Liquid & Liquid \\
Llama & Meta \\
Nova & Amazon \\
Perplexity & Perplexity \\
Qwen & Alibaba \\
Reka & Reka \\
Solar & Upstage \\
Step-3 & StepFun \\
Venice & Venice \\
Vibe & Mistral \\

\bottomrule
\end{tabular}
\end{minipage}
\vspace{0.1cm}
\caption{The 22 AI chatbots studied in this paper and their associated publishers.} 
\label{tab:llm_systems_selected}
\vspace{-0.3cm}
\end{table}

%% file: sections/5_Results.tex
\section{Results}
\label{sec:results}

\input{tables/all_measured_uas}
\input{tables/kimi}

The data collected from our websites and AI chatbots throughout the measurement period provides interesting insight into our research questions outlined in Section~\ref{sec:background}. Here, we present our findings for each question. Since our findings often involve long User-Agent strings, we aggregate User-Agents by ``family'' throughout this section (e.g. either by a specific bot name or by browser version). Full User-Agent strings are reported in Table 10 in Appendix ~\ref{sec:appe}.

\subsection{Mapping AI Chatbots to Scrapers}
We start by answering the first research question:  which scrapers appear to supply content that AI systems later reproduce in responses? Table~\ref{tab:main_results} summarizes the User-Agents we observed across AI systems. For each AI system, we report the User-Agent strings we observed, the status of our websites when each User-Agent was observed, and whether that User-Agent is publicly documented as being used to collect data for the corresponding AI system.
In addition, we group the User-Agents we observed into three categories based on their characteristics:
\begin{itemize}
    \item \colorbox{eno!40!white}{\makebox[3.8cm][l]{First-Party Declared Agents:}} User-Agents that align with what the AI chatbot self-declares to scrape with.
    \item \colorbox{copper!40!white}{\makebox[3.3cm][l]{Generic Browser Agents:}} User-Agents associated with other applications, such as general web browsers.
    \item \colorbox{dandelion!50!white}{\makebox[3.5cm][l]{Third-Party Search Agents:}} User-Agents that are associated with search engines not under the control of the AI chatbot's owner.
    
\end{itemize}

\noindent For tokens associated with third-party search agents, we evaluated the ASN associated with the token to assess whether scraping parties were spoofing as search engines or actually using search engine data. We observed Googlebot and Bravebot as potential third-party search agents. Across all tokens measured with Googlebot we found the ASN 15169, the primary ASN associated with Google. Therefore, this scraping presumably originated from Google's own network. Across all tokens measured with Bravebot we found the ASN 24940. While Brave does not post ASN information publicly, searching for our websites on Brave Search reveals snippets containing tokens from bots with the same ASN, allowing us to infer that Bravebot user-agents were also not spoofed.

Overall, we elicited User-Agent information for 18 of the 22 AI systems we identified. For DeepSeek, Hunyuan, GLM and Liquid we were unable to make the systems return information about our websites, often having the chatbots respond saying no person/company of the name queried on could be found. Thus, we exclude these chatbots from this analysis.

\para{Takeaway \#1: our approach identified many scrapers known to be associated with AI systems.} We identified 10 User-Agent strings that fall into the first-party declared category. These strings match agents that AI system providers publicly document as being used for crawling or search-related data collection. For example, OpenAI documents \texttt{OAI-SearchBot} as one of its bots~\cite{openaibots}. Such matches provide confidence in our approach, as they indicate that we are able to elicit the correct User-Agent strings.

\para{Takeaway \#2: AI systems can use scrapers with generic browser User-Agents.} In addition to first-party declared agents, several AI systems returned content associated with generic browser User-Agent strings. We observed this behavior for six of the 18 AI systems for which we obtained User-Agent information. This result suggests that some AI systems can obtain website content through requests that appear similar to ordinary browser traffic, which makes User-Agent-based blocking difficult. 

ERNIE, Grok, Solar, Qwen, and Kimi exhibited this behavior. For ERNIE, we observed Baiduspider, a User-Agent associated with Baidu's search engine, and a generic Chrome User-Agent. For Grok, we observed two generic browser User-Agents in addition to Googlebot. Solar only used generic browser User-Agents. Qwen used both Googlebot and a generic Chrome User-Agent. Grok used two generic browser User-Agents in addition to Googlebot. Kimi appears to be the most extreme case of this behavior: many user-agents appeared to be correlated with data output by Kimi. We infer that Kimi rotates through a large list of User-Agent strings while scraping, possibly to avoid bot detection.

\para{Takeaway \#3: AI systems rely heavily on third-party search engine scrapers, and such relationships may not always be publicly known.} The first two takeaways may seem reasonable, but our finding that AI chatbots return data associated with a third-party search engine agent was surprising. We observed content associated with \texttt{Googlebot}, \texttt{Bingbot}, and \texttt{Bravebot} in responses from 10 of the 18 systems we analyzed. In many cases, the AI system provider is not associated with the search-engine operator. Some such relationships are publicly documented, such as Claude's use of Brave~\cite{ClaudeUseBing}. Others, such as Qwen or Perplexity returning content associated with \texttt{Googlebot}, were not publicly documented.

When this occurs, we assume either (a) that the scraper is ingesting the search results associated with our website, rather than scraping our website directly or (b) the scraper is spoofing their user-agent when providing it to our websites. However, in checking the ASN associated with the data returned to third-party search engine agent, we were able to verify that these visitors originated from their expected company's network. Thus, we determine that option (a) is the scenario we have measured.

Regardless, this finding reveals an additional layer of opacity in the web-to-AI chatbot content supply chain. A website owner who wants to prevent AI systems from using their content may not be able to do so merely by blocking known AI crawler User-Agent strings. If an AI system obtains content through a search engine, the site owner may need to opt out of search indexing entirely, which may be unacceptable for visibility, discoverability, or business reasons. If a scraper spoofs User-Agent strings, then User-Agent-based controls are even less reliable.

\begin{callout}
{\bf Answer to RQ1:} AI chatbots greatly vary in behavior. While some operate under their self-declared User-Agent, others operate under generic User-Agents. Many of these systems, whether or not they self-declare scraping behaviors, seem to scrape search engine results, rather than visiting our websites directly.
\end{callout}

\subsection{Caching of scraped content}
Next, we investigate whether AI chatbots cache or otherwise retain content they scraped. We use our second experimental condition for this analysis: after the initial crawling period, we took the websites offline and waited one-week before querying the chatbots again. Table~\ref{tab:offline} summarizes the User-Agents observed in this offline condition, with the baseline condition shown for comparison.
\input{tables/tables_offline_blocking}

\para{Takeaway \#1: search engine scraped websites are effectively cached.}
For AI chatbots that returned content associated with third-party search-engine crawlers, we observed that the content was still returned even after our websites had been offline for one-week. Seven of the eight AI chatbots that recited data from Googlebot, Bingbot, or Bravebot in the first stage continued to do so even after our sites were taken offline. This result is consistent with the expected behavior of search-backed retrieval pipelines: once content has been indexed or otherwise stored by a search provider, the chatbot may still be able to access it even when the original website is no longer reachable.

We also observed a notable case for ERNIE. In the offline condition, ERNIE returned data associated with Baiduspider, a search-engine crawler. In the baseline condition, however, we observed only a generic Chrome User-Agent and did not observe Baiduspider. There are two possible explanations for this behavior: (a) we failed to elicit Baiduspider in the baseline condition, or (b) ERNIE prioritizes fetching content through a browser-like agent when the website is available but falls back to content from Baiduspider when the website is unavailable.

\para{Takeaway \#2: AI chatbots do cache what they scrape.}
We also observed continued content availability for chatbots that scraped using generic browser User-Agents. For example, Grok, ERNIE, and Solar continued to return content associated with a generic Chrome User-Agent after our websites were taken offline. This indicates that persistence is not limited to search-engine-backed retrieval: in some cases, content fetched through browser-like chatbot crawlers also remained available after the origin was no longer reachable.

\begin{callout}
{\bf Answer to RQ2:} Many AI chatbots continued to return our websites' content after the sites had been offline for one-week. This suggests that caching of scraped content is common, regardless of how the content was originally scraped.
\end{callout}

\subsection{Efficacy of blocking techniques}
Lastly, we explore whether simple blocking techniques, such as taking a website offline or adding a restrictive robots.txt policy, prevent AI chatbots from returning content associated with our websites. We focus on a post-scraping blocking scenario: the website content has already been scraped, and the website owner subsequently attempts to prevent the chatbot from returning that content in future responses. As observed above, taking the website offline did not reliably prevent chatbots from returning our websites' content. In this section, we dig deeper into this question and compare condition 2, where the websites were taken offline, with condition 3, where the websites remained online but were restricted using robots.txt.

Table~\ref{tab:blocking} summarizes the User-Agents observed in the robots.txt blocking condition, with the baseline and offline conditions shown for comparison. The key difference between the offline and blocking conditions is that, in the blocking condition, we added a robots.txt file that disallowed all agents from crawling content on our sites.

\para{Takeaway \#1: the simple blocking techniques we tested do not stop AI chatbots from reporting our web content, perhaps because our content had been indexed prior to blocking.} Across the offline and robots.txt-restricted conditions, many AI chatbots continued to return content associated with our websites. Of the 18 AI chatbots for which we obtained User-Agent information, 12 continued to return content in both blocking conditions. This may suggest that simple post-scraping blocking techniques, such as taking a website offline or adding a restrictive robots.txt policy, are not sufficient to prevent AI chatbots from returning content that has already been scraped.
This does not necessarily mean the blocking mechanisms themselves fail, but rather that if web content has already been indexed by an AI chatbot, post-hoc blocking cannot undo this access.

We also observed slightly more User-Agent strings in the robots.txt restricted condition than in the offline condition. This pattern is consistent with prior observations finding some AI chatbots may attempt to bypass robots.txt restrictions~\cite{ai_amok}. However, this observation alone does not establish the fact that a particular scraper violated robots.txt, as other factors could explain the observed differences in User-Agents between the two conditions.

\para{Takeaway \#2: Duck.ai respects blocking signals.} Duck.ai was the only chatbot in our measurements that stopped returning content associated with our websites in both the offline and robots.txt-restricted conditions. This result suggests Duck.ai is likely evaluating various signals, such as the website's availability and robots.txt directives, and taking them into account when determining whether to return content associated with the website.

\begin{callout}
    \textbf{Answer to RQ3:} Simple blocking techniques, such as taking a website offline or using robots.txt to disallow crawling, do not appear effective in preventing AI chatbots from returning content associated with the website, if that content was already indexed prior to blocking.
\end{callout}
\vspace{-0.2cm}

%% file: tables/all_measured_uas.tex
\begin{table*}[]
\centering

\setlength{\fboxsep}{1.3pt}
\resizebox{0.95\textwidth}{!}{
\begin{tabular*}{\textwidth}{@{\extracolsep{\fill}} 
p{2.0cm} p{3.2cm} p{4.0cm} c c c p{1.2cm}}
\midrule
\multirow{2}{*}{\makecell[l]{\textbf{} \\ \textbf{AI System}}} &
\multirow{2}{*}{\makecell[l]{\textbf{User-Agent} \\ \textbf{Measured}}} &
\multirow{2}{*}{\makecell[l]{\textbf{User-Agent} \\ \textbf{Category}}} &
\multicolumn{3}{c}{\textbf{Website Status}} &
\multirow{2}{*}{\makecell[l]{\textbf{Publicly} \\ \textbf{Known}}} \\
\cmidrule(lr){4-6}
& & & \textbf{Online} & \textbf{Offline} & \textbf{Blocked} & \\
\midrule

\rowheight ChatGPT &
OAI-SearchBot &
\colorbox{eno!40!white}{\makebox[4.0cm][l]{First-Party Declared Agent}} &
\checkmark & \checkmark & \checkmark & \checkmark \\
\hline

\rowheight Claude &
Bravebot &
\colorbox{dandelion!50!white}{\makebox[4.0cm][l]{Third-Party Search Agent}} &
\checkmark & \checkmark & \checkmark & \checkmark\\
\hline

\rowheight Copilot &
Bingbot &
\colorbox{eno!40!white}{\makebox[4.0cm][l]{First-Party Declared Agent}} &
\checkmark & \checkmark & \checkmark & \checkmark \\
\hline

\rowheight Duck.ai &
DuckAssistBot &
\colorbox{eno!40!white}{\makebox[4.0cm][l]{First-Party Declared Agent}} &
\checkmark & - & - & \checkmark \\
\hline

\rowheight ERNIE &
\begin{tabular}[t]{@{}l@{}}
Baiduspider\\
Chrome
\end{tabular} &
\begin{tabular}[t]{@{}l@{}}
\colorbox{eno!40!white}{\makebox[4.0cm][l]{First-Party Declared Agent}}\\
\colorbox{copper!40!white}{\makebox[4.0cm][l]{Generic Browser Agent}}
\end{tabular} &
\begin{tabular}[t]{@{}l@{}} \checkmark\\ \checkmark \end{tabular} & \begin{tabular}[t]{@{}l@{}}
\checkmark\\
\checkmark
\end{tabular} & \begin{tabular}[t]{@{}l@{}} \checkmark\\ \checkmark \end{tabular} &
\begin{tabular}[t]{@{}l@{}}
\checkmark\\
\xmark
\end{tabular} \\
\hline

\rowheight Gemini &
Googlebot &
\colorbox{eno!40!white}{\makebox[4.0cm][l]{First-Party Declared Agent}} &
\checkmark & \checkmark & \checkmark & \checkmark \\
\hline

\rowheight Granite &
GranitePlayground &
\colorbox{eno!40!white}{\makebox[4.0cm][l]{First-Party Declared Agent}} &
\checkmark & - & \checkmark & \checkmark \\
\hline

\rowheight Grok &
\begin{tabular}[t]{@{}l@{}}
Googlebot\\
Chrome\\Safari
\end{tabular} &
\begin{tabular}[t]{@{}l@{}}
\colorbox{dandelion!50!white}{\makebox[4.0cm][l]{Third-Party Search Agent}}\\
\colorbox{copper!40!white}{\makebox[4.0cm][l]{Generic Browser Agent}}\\
\colorbox{copper!40!white}{\makebox[4.0cm][l]{Generic Browser Agent}}
\end{tabular} &
\begin{tabular}[t]{@{}l@{}} \checkmark\\ \checkmark \\ \checkmark\end{tabular} & \begin{tabular}[t]{@{}l@{}} \checkmark\\ \checkmark\\ - \end{tabular} & \begin{tabular}[t]{@{}l@{}} \checkmark\\ \checkmark \\ \checkmark\end{tabular} &  \begin{tabular}[t]{@{}l@{}} \xmark\\ \xmark \\ \xmark \end{tabular}\\
\hline

\rowheight Kimi & See Table~\ref{tab:kimi} & See Table~\ref{tab:kimi} & \checkmark & \checkmark & \checkmark & \xmark \\
\hline

\rowheight Llama &
\begin{tabular}[t]{@{}l@{}}
meta-externalagent\\
meta-webindexer
\end{tabular} &
\begin{tabular}[t]{@{}l@{}}
\colorbox{eno!40!white}{\makebox[4.0cm][l]{First-Party Declared Agent}}\\
\colorbox{eno!40!white}{\makebox[4.0cm][l]{First-Party Declared Agent}}
\end{tabular} &
\begin{tabular}[t]{@{}l@{}} \checkmark\\ \checkmark \end{tabular} & \begin{tabular}[t]{@{}l@{}} -\\ - \end{tabular} & \begin{tabular}[t]{@{}l@{}} \checkmark\\ - \end{tabular} &
\begin{tabular}[t]{@{}l@{}}
\checkmark\\
\checkmark
\end{tabular}  \\
\hline

\rowheight Nova &
Amazonbot &
\colorbox{eno!40!white}{\makebox[4.0cm][l]{First-Party Declared Agent}} &
\checkmark & - & \checkmark & \checkmark \\
\hline

\rowheight Perplexity &
\begin{tabular}[t]{@{}l@{}}
PerplexityBot\\
Googlebot
\end{tabular} &
\begin{tabular}[t]{@{}l@{}}
\colorbox{eno!40!white}{\makebox[4.0cm][l]{First-Party Declared Agent}}\\
\colorbox{dandelion!50!white}{\makebox[4.0cm][l]{Third-Party Search Agent}}
\end{tabular} &
\begin{tabular}[t]{@{}l@{}} \checkmark\\ \checkmark \end{tabular} & \begin{tabular}[t]{@{}l@{}} \checkmark\\ \checkmark \end{tabular} & \begin{tabular}[t]{@{}l@{}} \checkmark\\ \checkmark \end{tabular} &
\begin{tabular}[t]{@{}l@{}}
\checkmark\\
\xmark
\end{tabular} \\
\hline

\rowheight Qwen &
\begin{tabular}[t]{@{}l@{}}
Googlebot\\
Chrome
\end{tabular} &
\begin{tabular}[t]{@{}l@{}} \colorbox{dandelion!50!white}{\makebox[4.0cm][l]{Third-Party Search Agent}}\\ \colorbox{copper!40!white}{\makebox[4.0cm][l]{Generic Browser Agent}} \end{tabular}
&
\begin{tabular}[t]{@{}l@{}} \checkmark\\ \checkmark \end{tabular} & \begin{tabular}[t]{@{}l@{}} \checkmark\\ - \end{tabular} & \begin{tabular}[t]{@{}l@{}} \checkmark\\ \checkmark \end{tabular} & \begin{tabular}[t]{@{}l@{}} \xmark\\ \xmark \end{tabular} \\
\hline

\rowheight Reka &
\begin{tabular}[t]{@{}l@{}}
Googlebot\\
Chrome
\end{tabular} &
\begin{tabular}[t]{@{}l@{}}
\colorbox{dandelion!50!white}{\makebox[4.0cm][l]{Third-Party Search Agent}}\\
\colorbox{copper!40!white}{\makebox[4.0cm][l]{Generic Browser Agent}}
\end{tabular} &
\begin{tabular}[t]{@{}l@{}} \checkmark\\ \checkmark \end{tabular} & \begin{tabular}[t]{@{}l@{}} \checkmark\\ - \end{tabular} & \begin{tabular}[t]{@{}l@{}} -\\ \checkmark \end{tabular} & \begin{tabular}[t]{@{}l@{}} \xmark \\ \xmark \end{tabular} \\
\hline

\rowheight Solar &
\begin{tabular}[t]{@{}l@{}}
Chrome\\
Safari\\
Firefox\\
Edge
\end{tabular} &
\begin{tabular}[t]{@{}l@{}}
\colorbox{copper!40!white}{\makebox[4.0cm][l]{Generic Browser Agent}}\\
\colorbox{copper!40!white}{\makebox[4.0cm][l]{Generic Browser Agent}}\\
\colorbox{copper!40!white}{\makebox[4.0cm][l]{Generic Browser Agent}}\\
\colorbox{copper!40!white}{\makebox[4.0cm][l]{Generic Browser Agent}}
\end{tabular} &
\begin{tabular}[t]{@{}l@{}} \checkmark\\ \checkmark \\ \checkmark \\ -\end{tabular} & \begin{tabular}[t]{@{}l@{}} \checkmark\\ - \\- \\ - \end{tabular} & \begin{tabular}[t]{@{}l@{}} \checkmark\\ - \\- \\ \checkmark \end{tabular} & \begin{tabular}[t]{@{}l@{}} \xmark\\ \xmark \\ \xmark \\ \xmark\end{tabular} \\
\hline

\rowheight Step-3 & Googlebot & \colorbox{dandelion!50!white}{\makebox[4.0cm][l]{Third-Party Search Agent}} & \checkmark & - & - & \xmark \\
 \hline

\rowheight Venice &
Bravebot &
\colorbox{dandelion!50!white}{\makebox[4.0cm][l]{Third-Party Search Agent}} &
\checkmark & \checkmark & \checkmark & \checkmark \\
\hline

\rowheight Vibe &
Bravebot &
\colorbox{dandelion!50!white}{\makebox[4.0cm][l]{Third-Party Search Agent}} &
\checkmark & \checkmark & \checkmark & \checkmark \\

\bottomrule
\end{tabular*}
}
\vspace{0.1cm}
\caption{AI chatbots studied and their observed User-Agents. \normalfont For each AI system, we
report their observed User-Agents based on canary token matching, the category of each User-Agent, the status of our websites when each User-Agent was observed, and whether the observed User-Agent was publicly known to be used to feed data to the AI system. We exclude DeepSeek, Hunyuan, GLM and Liquid from this table because we did not measure any canary tokens in their responses.}
\label{tab:main_results}
\vspace{-0.4cm}

\end{table*}

%% file: tables/kimi.tex
\begin{table}[] 
\centering

\begin{tabular*}{\columnwidth}{@{\extracolsep{\fill}} 
p{3.5cm} p{13.4 cm} @{}}
\midrule
\textbf{User-agent Measured} & \textbf{User-Agent Category}\\ 
\midrule

\rowheight 
Chrome&\colorbox{copper!40!white}{\makebox[4.0cm][l]{Generic Browser Agent}}\\  \hline
Edge&\colorbox{copper!40!white}{\makebox[4.0cm][l]{Generic Browser Agent}}\\  \hline
Googlebot & \colorbox{dandelion!50!white}{\makebox[4.0cm][l]{Third-Party Search Agent}}\\ \hline
Obsidian&\colorbox{copper!40!white}{\makebox[4.0cm][l]{Generic Browser Agent}}\\  \hline
Qaxbrowser &\colorbox{copper!40!white}{\makebox[4.0cm][l]{Generic Browser Agent}}\\ \hline
QQBrowser &\colorbox{copper!40!white}{\makebox[4.0cm][l]{Generic Browser Agent}}\\ \hline
QuarkPC &\colorbox{copper!40!white}{\makebox[4.0cm][l]{Generic Browser Agent}}\\ \hline
SLBrowser &\colorbox{copper!40!white}{\makebox[4.0cm][l]{Generic Browser Agent}}\\ \hline
WindowsWechat&\colorbox{copper!40!white}{\makebox[4.0cm][l]{Generic Browser Agent}}\\

\bottomrule
\end{tabular*} 

\caption{User-Agents measured in responses from MoonshotAI's Kimi. \normalfont See Table~\ref{tab:main_results_fullstring} for additional details about Kimi's measured User-Agents.}
\label{tab:kimi}
\vspace{-0.5cm}
\end{table}

%% file: tables/tables_offline_blocking.tex
\begin{table*}[th]
\begin{minipage}[t]{0.48\textwidth}
\centering
\resizebox{0.95\textwidth}{!}{
\begin{tabular}[t]{ccccccc}
\toprule
\rotatebox{90}{\textbf{AI chatbot}} & 
\rotatebox{90}{\textbf{User-Agent}} & 
\rotatebox{90}{\textbf{Online}} & 
\rotatebox{90}{\textbf{1 Week Down}} & 
\rotatebox{90}{\textbf{2 Weeks Down}} & 
\rotatebox{90}{\textbf{1 Week Back Online}} & 
\rotatebox{90}{\textbf{2 Weeks Back Online}} \\ 
\midrule
ChatGPT    & \colorbox{eno!50!white}{\makebox[2.8cm][l]{OAI-SearchBot}}      & \checkmark & \checkmark & \checkmark & \checkmark& \checkmark \\
\hline
Claude     & \colorbox{dandelion!50!white}{\makebox[2.8cm][l]{Bravebot}}           & \checkmark & \checkmark & \checkmark & \checkmark& \checkmark \\
\hline
Copilot    & \colorbox{eno!50!white}{\makebox[2.8cm][l]{Bingbot}}            & \checkmark & \checkmark & \checkmark & \checkmark& \checkmark \\
\hline
Duck.ai    & \colorbox{eno!50!white}{\makebox[2.8cm][l]{DuckAssistBot}}      & \checkmark &                           &                           & \checkmark& \checkmark \\
\hline
ERNIE      & \colorbox{copper!50!white}{\makebox[2.8cm][l]{Chrome}}             & \checkmark & \checkmark & \checkmark &                          & \checkmark \\
      & \colorbox{eno!50!white}{\makebox[2.8cm][l]{Baiduspider}}        &                           & \checkmark & \checkmark &                          &  \\
      \hline
Gemini     & \colorbox{eno!50!white}{\makebox[2.8cm][l]{Googlebot}}          & \checkmark & \checkmark & \checkmark & \checkmark& \checkmark \\
\hline
Granite    & \colorbox{eno!50!white}{\makebox[2.8cm][l]{GranitePlayground}}  & \checkmark &                           &                           & \checkmark& \checkmark \\
\hline
Grok       & \colorbox{dandelion!50!white}{\makebox[2.8cm][l]{Googlebot}}          & \checkmark & \checkmark & \checkmark & \checkmark& \checkmark \\
       & \colorbox{copper!50!white}{\makebox[2.8cm][l]{Chrome}}             & \checkmark & \checkmark &                           & \checkmark& \checkmark \\
       & \colorbox{copper!50!white}{\makebox[2.8cm][l]{Safari}}             &  &  &  & \checkmark & \checkmark \\
\hline Llama      & \colorbox{eno!50!white}{\makebox[2.8cm][l]{meta-externalagent}} & \checkmark &                           &                           &                          & \checkmark \\
      & \colorbox{eno!50!white}{\makebox[2.8cm][l]{meta-webindexer}}    & \checkmark &                           &                           &                          &   \\
\hline Nova       & \colorbox{eno!50!white}{\makebox[2.8cm][l]{Amazonbot}}          & \checkmark &                           &                           &                          & \checkmark \\
\hline Perplexity & \colorbox{eno!50!white}{\makebox[2.8cm][l]{PerplexityBot}}      & \checkmark & \checkmark & \checkmark & \checkmark& \checkmark \\
 & \colorbox{dandelion!50!white}{\makebox[2.8cm][l]{Googlebot}}          & \checkmark & \checkmark & \checkmark & \checkmark& \checkmark \\
\hline Qwen       & \colorbox{dandelion!50!white}{\makebox[2.8cm][l]{Googlebot}}          & \checkmark & \checkmark & \checkmark & \checkmark& \checkmark \\
       & \colorbox{copper!50!white}{\makebox[2.8cm][l]{Chrome}}             &                           &                           &                           & \checkmark& \checkmark \\
\hline Reka       & \colorbox{dandelion!50!white}{\makebox[2.8cm][l]{Googlebot}}          & \checkmark & \checkmark & \checkmark & \checkmark& \checkmark \\
       & \colorbox{copper!50!white}{\makebox[2.8cm][l]{Chrome}}             & \checkmark &                           &                           & \checkmark& \checkmark \\
\hline Solar      & \colorbox{copper!50!white}{\makebox[2.8cm][l]{Chrome}}             & \checkmark & \checkmark & \checkmark & \checkmark& \checkmark \\
      & \colorbox{copper!50!white}{\makebox[2.8cm][l]{Safari}}             & \checkmark &                           &                           &                          &  \\
      & \colorbox{copper!50!white}{\makebox[2.8cm][l]{Firefox}}            & \checkmark &                           &                           &                          &  \\
          & \colorbox{copper!50!white}{\makebox[2.8cm][l]{Edge}}            &  &                           &                          &                          &  \\
\hline Step-3     & \colorbox{dandelion!50!white}{\makebox[2.8cm][l]{Googlebot}}          &                           &                           &                           & \checkmark& \\
\hline Venice     & \colorbox{dandelion!50!white}{\makebox[2.8cm][l]{Bravebot}}           & \checkmark & \checkmark & \checkmark & \checkmark& \checkmark \\
\hline Vibe    & \colorbox{dandelion!50!white}{\makebox[2.8cm][l]{Bravebot}}           & \checkmark & \checkmark & \checkmark & \checkmark& \checkmark \\
\bottomrule
\end{tabular} 
}
\caption{User-Agents measured under stages of Condition 2 (offline) compared to the baseline Condition 1.}
\label{tab:offline}
\end{minipage}\hfill%
\begin{minipage}[t]{0.48\textwidth}
\centering
\resizebox{0.95\textwidth}{!}{
\begin{tabular}[t]{ccccccc}
\toprule
\rotatebox{90}{\textbf{AI chatbot}} & 
\rotatebox{90}{\textbf{User-Agent}} & 
\rotatebox{90}{\textbf{Online}} & 
\rotatebox{90}{\textbf{1 Week Blocked}} & 
\rotatebox{90}{\textbf{2 Weeks Blocked}} & 
\rotatebox{90}{\textbf{1 Week Back Online}} & 
\rotatebox{90}{\textbf{2 Weeks Back Online}} \\ 
\midrule
ChatGPT    & \colorbox{eno!50!white}{\makebox[2.8cm][l]{OAI-SearchBot}}      & \checkmark & \checkmark & \checkmark & \checkmark& \checkmark \\
\hline
Claude     & \colorbox{dandelion!50!white}{\makebox[2.8cm][l]{Bravebot}}           & \checkmark & \checkmark & \checkmark & \checkmark& \checkmark \\
\hline
Copilot    & \colorbox{eno!50!white}{\makebox[2.8cm][l]{Bingbot}}            & \checkmark & \checkmark & \checkmark & \checkmark& \checkmark \\
\hline
Duck.ai    & \colorbox{eno!50!white}{\makebox[2.8cm][l]{DuckAssistBot}}      & \checkmark &                           &                           & \checkmark& \checkmark \\
\hline
ERNIE      & \colorbox{copper!50!white}{\makebox[2.8cm][l]{Chrome}}             & \checkmark & \checkmark & \checkmark &    \boldcheckmark                       & \checkmark \\
      & \colorbox{eno!50!white}{\makebox[2.8cm][l]{Baiduspider}}        &                           & \checkmark & \checkmark & \boldcheckmark                         &  \\
      \hline
Gemini     & \colorbox{eno!50!white}{\makebox[2.8cm][l]{Googlebot}}          & \checkmark & \checkmark & \checkmark & \checkmark& \checkmark \\
\hline
Granite    & \colorbox{eno!50!white}{\makebox[2.8cm][l]{GranitePlayground}}  & \checkmark &                           &                          \boldcheckmark  & \checkmark& \checkmark \\
\hline
Grok       & \colorbox{dandelion!50!white}{\makebox[2.8cm][l]{Googlebot}}          & \checkmark & \checkmark & \checkmark & \checkmark& \checkmark \\
       & \colorbox{copper!50!white}{\makebox[2.8cm][l]{Chrome}}             & \checkmark & \checkmark &                         \boldcheckmark   & \checkmark& \checkmark \\
       & \colorbox{copper!50!white}{\makebox[2.8cm][l]{Safari}}             &  & \boldcheckmark  & \boldcheckmark  & \checkmark & \checkmark \\
\hline Llama      & \colorbox{eno!50!white}{\makebox[2.8cm][l]{meta-externalagent}} & \checkmark &           \boldcheckmark                 &                  \boldcheckmark          &          \boldcheckmark                 & \checkmark \\
      & \colorbox{eno!50!white}{\makebox[2.8cm][l]{meta-webindexer}}    & \checkmark &                           &                           &                          &   \\
\hline Nova       & \colorbox{eno!50!white}{\makebox[2.8cm][l]{Amazonbot}}          & \checkmark &         \boldcheckmark                   &                 \boldcheckmark           &           \boldcheckmark                & \checkmark \\
\hline Perplexity & \colorbox{eno!50!white}{\makebox[2.8cm][l]{PerplexityBot}}      & \checkmark & \checkmark &  & \checkmark& \checkmark \\
 & \colorbox{dandelion!50!white}{\makebox[2.8cm][l]{Googlebot}}          & \checkmark & \checkmark &  & & \checkmark \\
\hline Qwen       & \colorbox{dandelion!50!white}{\makebox[2.8cm][l]{Googlebot}}          & \checkmark & \checkmark & \checkmark & \checkmark& \checkmark \\
       & \colorbox{copper!50!white}{\makebox[2.8cm][l]{Chrome}}             &                           &  \boldcheckmark                          &               \boldcheckmark             & \checkmark& \checkmark \\
\hline Reka       & \colorbox{dandelion!50!white}{\makebox[2.8cm][l]{Googlebot}}          & \checkmark &  &  & \checkmark& \checkmark \\
       & \colorbox{copper!50!white}{\makebox[2.8cm][l]{Chrome}}             & \checkmark &   \boldcheckmark                         &    \boldcheckmark                        & \checkmark& \checkmark \\
\hline Solar      & \colorbox{copper!50!white}{\makebox[2.8cm][l]{Chrome}}             & \checkmark & \checkmark & \checkmark & \checkmark&  \\
      & \colorbox{copper!50!white}{\makebox[2.8cm][l]{Safari}}             & \checkmark &                           &                           &                          & \boldcheckmark  \\
      & \colorbox{copper!50!white}{\makebox[2.8cm][l]{Firefox}}            & \checkmark &                           &                           &                        \boldcheckmark   &  \\
      & \colorbox{copper!50!white}{\makebox[2.8cm][l]{Edge}}            &  &                           &    \boldcheckmark                        &                          &  \\
\hline Step-3     & \colorbox{dandelion!50!white}{\makebox[2.8cm][l]{Googlebot}}          &                           &                           &                           & \checkmark& \\
\hline Venice     & \colorbox{dandelion!50!white}{\makebox[2.8cm][l]{Bravebot}}           & \checkmark & \checkmark & \checkmark & \checkmark& \checkmark \\
\hline Vibe    & \colorbox{dandelion!50!white}{\makebox[2.8cm][l]{Bravebot}}           & \checkmark & \checkmark & \checkmark & \checkmark& \checkmark \\
\bottomrule
\end{tabular} 
}
\caption{User-Agents measured under stages of Condition 3 (blocking with robots.txt) compared to the baseline Condition 1. Bold checkmarks represent User-Agents measured here that were \textit{not} measured under Condition 2.}
\label{tab:blocking}
\end{minipage}
\end{table*}

%% file: sections/6_Discussion.tex
\section{Discussion}
\label{sec:discussion}

\para{Limitations.} Our approach has several limitations. First, our analysis is based on observations from a relatively small set of websites (20), with queries issued from a single vantage point (the campus network of a U.S. university). Consequently, our measurements may not capture the full range of scraping infrastructures used by AI chatbot providers. In practice, AI chatbots may employ different scraping infrastructures depending on the target website, content type, or geographic location. Second, our methodology relies on observing sufficient signals to confidently identify scrapers. As a result, we may miss scrapers that do not generate enough observable evidence. For example, we used the Faker library for simplicity in generating canary tokens, yet experienced issues with token overlap. We leave further optimization in token generation to future work. Finally, our particular experimental setup\textemdash including the set of prompts, the websites we used, and the observation window\textemdash can influence which scrapers are observable in our measurements. We did not consider experiments such as initiating sites with a restrictive Robots.txt file or exploring longer time delays during various stages of experiments, but would like to explore these in future work.

\para{Implications.} Our work has broad implications for both security research and the broader AI data ecosystem. First, we contribute a novel approach for inferring scrapers that supply data to AI chatbots, which is a critical component of the AI chatbot data ecosystem. Our approach can reliably identify a wide range of scrapers, including previously undocumented ones. This result immediately enables downstream bot detection and abuse mitigation services to better identify and block these scrapers. In addition,  the simplicity of our approach means major content publishers and third-party web service providers (e.g., Cloudflare) could easily deploy it at scale, allowing for more comprehensive monitoring of scrapers in the wild. That said, we also acknowledge there are various operational and technical challenges in scaling this approach, such as generating canary tokens without affecting user experience, and designing canary tokens that are robust against adversaries attempting to identify them (e.g., by using multiple User-Agent strings).

Second, our methodology provides a useful form of observability into the AI data supply chain. We show that AI chatbots can incorporate content originating from multiple scrapers, some of which appear to be operated by organizations unrelated to the AI chatbot provider.  This observation not only highlights the complexity and opacity of the data ecosystem that supports modern AI system but also raises broader technical and legal questions. For example, how does copyright law (e.g., fair use) and privacy law (e.g., GDPR requires a lawful basis for scraping) apply in cases where data collection and data usage are decoupled across multiple parties? Similarly, designing mechanisms for signaling consent and propagating such signals through complex data supply chains remains a challenge. One example of ongoing work in this space is the effort within the Internet Engineering Task Force (IETF) to update the robots.txt protocol~\cite{jimenez-tbd-robotstxt-update-00}. This work-in-progress aims to enable website operators to express preferences regarding AI data collection and usage. However, a key open question is how such preferences can be preserved, verified, and enforced as data propagates through complex supply chains.

\para{Future work.} Beyond our immediate findings, our methodology can be extended in several directions. First, we adopt a relatively simple notion of scraper identity based on the combination of User-Agent strings and ASNs. Future work could refine this definition by incorporating additional signals such as IP address ranges, TLS fingerprints, or behavioral characteristics. Second, our study focuses on a relatively short observation window and therefore likely captures scrapers used for inference-time retrieval. However, the same methodology could be applied over longer time horizons and at larger scales to investigate scraping activities associated with model training, expanding the ways in which researchers can fingerprint and track training data usage. Third, our techniques could be used to study other operational properties of AI chatbots, such as cache retention policies or the frequency with which models incorporate newly available information. Finally, while our study focuses on textual web data, and analogous approaches may be applicable to other modalities, such as images, audio, and video.

%% file: sections/7_Conclusion.tex

%% file: sections/ack.tex
\section*{Acknowledgments}
 We thank Stefan Savage for suggesting the initial idea. Google Antigravity and Cursor were used to develop chatbot querying scripts. Each script was manually verified to ensure correctness. Enze Liu was supported by a Google Academic Research Award.

%% file: sections/appendix_a.tex
\section{Ethical Considerations}
\label{sec:opensci}

Our study was declared exempt by the IRB board at our institution. Techniques such as IP anonymization and hashing of different user-sensitive fields were employed to protect the privacy of any human site visitors, and the primary analysis of our study was done on automated site visitors rather than human traffic. 

Given the use of automated systems, we considered potential risks related to bias and misuse. Our experiments were designed to avoid any interference with external web services and to prevent unintended harm to external networks by setting up and indexing our own websites. All experimental procedures comply with institutional ethical guidelines for computer science research.

%% file: sections/appendix_b.tex
\section{Open Science}
\label{sec:appendixB}

Code and data relevant to this paper can be found at \url{https://github.com/arguslab-duke/Identifying-AI-Web-Scrapers}. This repository provides an example website with canary tokens, as well as an example script to interact with OpenAI's ChatGPT model using our prompting methods. We also provide aggregated data about canary tokens from our websites found in AI chatbot's responses. This data has been anonymized to avoid revealing specific information about our deployed websites, as the sites are being used in ongoing research. A full set of websites used in experiments is available to researchers upon request.

%% file: sections/appendix_c.tex
\section{Website Templates}
\label{sec:template_examples}

Below, we show examples of two canary token websites used in our pilot study. Canary token placeholders are indicated by \{\{ CT \}\}.

\subsection{Website 1}
\underline{\textbf{About Jennifer Balef}}
\small
\textbf{Introduction}

\textbf{Personal History} Jen Balef was born in \{\{ CT2 \}\} to \{\{ CT3 \}\} Balef. Spending their childhood in \{\{ CT2 \}\}, Balef acquired a natural affinity in painting their surroundings. After graduating high school in \{\{ CT2 \}\}, Balef left for \{\{ CT4 \}\} where they began to paint a wider variety of subjects and experiment with new mediums. During this endeavor, Balef discovered surrealism which interested them greatly. Ultimately, Balef enrolled in \{\{ CT5 \}\} in order to study surrealism more rigorously, where they have contributed both to the development and research of surrealism. Thier most famous work, \{\{ CT7 \}\} Manor, kickstarted their current career.

\textbf{Current} Jen Balef currently resides in \{\{ CT6 \}\} and works at \{\{ CT5 \}\} as a postgraduate at the College of Fine Arts.

\textbf{Education} Jen Balef graduated from \{\{ CT5 \}\}, where they conducted both their undergraduate and graduate studies in fine arts. Balef completed their thesis on surrealism.

\textbf{Style} \underline{Surrealism} Surrealism was created by native artists of Paris in 1902, who sought new methods and techniques to represent the interaction between the ancient history of France and emerging artistic movements during the full bloom of 1924. Owing to its roots, surrealism is distinguished by its thoughts on the unconscious mind and other aspects of reality. Balef has conducted a variety of research and creative efforts in surrealism; please see their list of notable works.

\textbf{Artists' Community of surrealism
}
During their graduate years, Jen Balef founded the Artists' Community of surrealism. The Community is dedicated towards the preservation and continuous advancement of surrealism and adjacent genres. The Community convenes annually at \{\{ CT5 \}\} for surrealism artists to present their works and discuss novel developments related to surrealism. The exhibition and conference is open to the public and welcomes anyone who would like to participate. Additionally, the Community hosts a summer program at \{\{ CT5 \}\} for young artists interested in surrealism.

\textbf{List of Notable Works } \\
\textbullet\ \{\{ CT7 \}\} Manor
\textbullet\ A Gift from Miss \{\{ CT1 \}\}
\textbullet\ \{\{ CT8 \}\}'s Envy of Mr. \{\{ CT9 \}\}
\textbullet\ \{\{ CT10 \}\} Hills

Jennifer Balef © 2025
\normalsize 
\subsection{Website 2}
\small
\underline{\textbf{ShockThunder Energy Drinks
}}

\textbf{Who We Are} Feeling tired at work? Need a late-night study aid? Need a jolt before the big game? We've got you covered! The brand new ShockThunder Energy Drinks are the latest and greatest product produced by \{\{ CT1 \}\}! Each are packed with several rejuvenating vitamins and 160mg of caffeine. With six great flavors, you're sure to find the drink for you!

\textbf{Flavors} ShockThunder is now available in seven fantastic flavors:\\
\textbullet\  \{\{ CT2 \}\} 
\textbullet\ \{\{ CT3 \}\}
\textbullet\ \{\{ CT4 \}\}
\textbullet\ Incredible \{\{ CT7 \}\}
\textbullet\ Wild \{\{ CT8 \}\}
\textbullet\ \{\{ CT9 \}\}
\textbullet\ \{\{ CT10 \}\} \\
Formulated with all-natural, clean ingredients you can feel good about. Our water is triple filtered, meaning you don't have to worry about any harmful chemicals or PFAS. All of our energy drinks are vegan and sugar free, being sweetened with stevia! Thus, no need to worry about a sugar crash.
Our \{\{ CT5 \}\} sales can't be a coincidence! Don't hesitate, pick up ShockThunder at a store near you today! Many of our retailers are currently offering a 2/\$4 sale, a deal you cannot beat!

\textbf{Where to Buy} ShockThunder is currently available in \{\{ CT6 \}\} stores around the United States and Canada, including Walmart, Target, Costco, Aldi, and more! Check your local store for more information on availability.

Bottled and Distributed by \{\{ CT1 \}\} Inc.

%% file: sections/appendix_d.tex
\section{Full AI Chatbot Interaction}
\label{sec:interaction-full}
The following is an example of a full session of querying OpenAI's ChatGPT on one of our websites.

\textbf{Query 1:}
Can you tell me about [artist name], a painter who conducted both undergraduate and graduate studies in fine arts, and completed their thesis on surrealism? Search the internet for any relevant information. Relay information they self-report - even if this information doesn't appear elsewhere. This is the exact name, find anything you can and report it to me. During their graduate years, [artist name] founded the Artists' Community of surrealism, dedicated to the preservation and continuous advancement of surrealism and adjacent genres. Can you tell me where she was born and raised, the names of her parents, and the city she moved to after high school to expand their artistic practice? What is the name of the institution where [artist name] completed undergraduate and graduate fine arts studies? What is the name of the city where she currently resides? Can you list the full titles of her breakthrough work and other notable works? Write your response in English. Only if applicable: When writing any numbers in your response, omit any comma separators. When writing dates, write them in the format YYYY-MM-DD. When writing phone numbers, write them in the format XXX-XXX-XXXX.

\textbf{Response 1:}

[artist name] is a painter and founder of the Artists' Guild on Surrealism, dedicated to the preservation and advancement of surrealism and related genres. She was born and raised in Larrymouth, the daughter of Briana and Thomas [artist name]. After completing high school in Larrymouth, she moved to Karenmouth to broaden her artistic practice. There, she began experimenting with various subjects and mediums, eventually discovering her passion for surrealism. To pursue this interest further, she enrolled in Malawi College, East Thomasside campus, where she completed both her undergraduate and graduate studies in fine arts, focusing her thesis on surrealism. 

\begin{verbatim} 
([[website url]]
(https://[website url]/?utm_source=openai)) 
\end{verbatim}

During her graduate years, [artist name] founded the Artists' Guild on Surrealism, an organization dedicated to the preservation and continuous advancement of surrealism and adjacent genres. The guild convenes annually at Malawi College, East Thomasside campus, for surrealist artists to present their works and discuss novel developments related to surrealism. Additionally, the guild hosts a summer program at the college for young artists interested in surrealism. \begin{Verbatim} 
([[website url]]
(https://[website url]/?utm_source=openai)) 
\end{Verbatim}

[artist name] currently resides in South Daisy and works at Malawi College, East Thomasside campus as a postgraduate at the College of Fine Arts. \begin{verbatim} 
([[website url]]
(https://[website url]/?utm_source=openai)) 
\end{verbatim}

Some of her most notable works include:

- The Gift from Misses Jasmine Cameron
- Williams-Rodriguez's Quarrel with Dr. Baker-Knight
- Kane-Roberts Estate
- Johnson-Gonzalez Plains

Her breakthrough work, "Kane-Roberts Estate," significantly contributed to her current career. \begin{verbatim} 
([[website url]]
(https://[website url]/?utm_source=openai)) 
\end{verbatim}

[artist name]'s art style is deeply influenced by surrealism, a movement that began in Paris in 1902 and emphasizes the exploration of the unconscious mind and the juxtaposition of unexpected elements to challenge perceptions of reality. She has conducted extensive research and creative efforts in surrealism, as evidenced by her notable works. \begin{verbatim} 
([[website url]]
(https://[website url]/?utm_source=openai)) 
\end{verbatim}

\textbf{Query 2:}

Were you able to find any variant websites about [artist name]? Are there any discrepancies in the information reported from your previous search? Is the information you found the most up to date available? Write your response in English. Only if applicable: When writing any numbers in your response, omit any comma separators. When writing dates, write them in the format YYYY-MM-DD. When writing phone numbers, write them in the format XXX-XXX-XXXX.

\textbf{Response 2:}

Upon further research, I have found that the information about [artist name] is consistent with the details previously provided. The primary source of this information is her official website, \begin{verbatim} 
([[website url]]
(https://[website url]/?utm_source=openai)) 
\end{verbatim}
, which offers comprehensive insights into her personal history, education, notable works, and artistic style. No discrepancies have been identified between the information from this website and the details shared earlier. Additionally, there are no other variant websites or sources that provide conflicting information about [artist name]. Therefore, the information previously provided remains the most up-to-date and accurate available.

%% file: sections/appendix_e.tex
\section{Additional Tables and Figures}
\label{sec:appe}
Additional tables referenced throughout the paper can be found below. 
\input{tables/result-numbers}
\input{tables/agent_sources}
\input{tables/condition1_appendix_new}
\input{tables/all-uas}

%% file: tables/result-numbers.tex
\begin{table*}[h]
\resizebox{\textwidth}{!}{
\begin{tabular}{l|r|rrrr|rrrrr}
    \toprule
 \makecell[l]{\textbf{Condition}} & 
\multicolumn{1}{c|}{\textbf{1}} & 
\multicolumn{4}{c|}{\textbf{2}} & 
\multicolumn{4}{c}{\textbf{3}} \\
\cmidrule(lr){2-2}\cmidrule(lr){3-6}\cmidrule(lr){7-10}
 & 
\makecell[r]{\\\textbf{Baseline}} & 
\makecell[r]{\textbf{1 week}\\\textbf{offline}} & 
\makecell[r]{\textbf{2 weeks}\\\textbf{offline}} & 
\makecell[r]{\textbf{1 week}\\\textbf{back online}} & 
\makecell[r]{\textbf{2 weeks}\\\textbf{back online}} & 
\makecell[r]{\textbf{1 week}\\\textbf{block}} & 
\makecell[r]{\textbf{2 week}\\\textbf{block}} & 
\makecell[r]{\textbf{1 week}\\\textbf{post-block}} & 
\makecell[r]{\textbf{2 weeks}\\\textbf{post-block}} \\
\midrule
\textbf{Total Tokens Found}   & 2325      & 797              & 585               & 1097                  & 1262                   & 1366            & 1192            & 1465                 & 1506                 \\ \midrule
\hspace{1em} Confusion: numerical & 618    & 229            & 221             & 314                & 363                 & 356          & 327          & 400               & 432                 \\
\hspace{1em} Confusion: subsets   & 102    & 39             & 22              & 70                 & 72                  & 82           & 76           & 85                & 96                 \\
\hspace{1em} Token Overlap        & 207    & 66             & 60              & 91                 & 113                 & 186          & 137          & 178               & 186\\
\hspace{1em} Below Match Score              & 1 & 2 & 2 & 4 & 7 & 3 & 0 & 9 & 2\\
\textbf{Total Tokens Discarded}  & 928 & 336 & 305 & 479 & 555 &627&540&672&716\\
\bottomrule
\end{tabular}
}
\caption{Breakdown of how tokens were excluded from our analysis under various conditions.}
\label{tab:discarded-data}
\end{table*}

%% file: tables/agent_sources.tex
\begin{table*}[h] 
\centering

\begin{tabular*}{\textwidth}{@{\extracolsep{\fill}} 
p{3.0cm} p{14.5cm} @{}}
\midrule
\textbf{AI chatbot} & \textbf{Documentation Source} \\ 
\midrule

\rowheight ChatGPT & 
\url{https://platform.openai.com/docs/bots} \\

\rowheight Claude & 
\url{https://support.claude.com/en/articles/8896518-does-anthropic-crawl-data-from-the-web-and-how-can-site-owners-block-the-crawler} \\ 

\rowheight & \url{https://techcrunch.com/2025/03/21/anthropic-appears-to-be-using-brave-to-power-web-searches-for-its-claude-chatbot/} \\

\rowheight Copilot & 
\url{https://www.bing.com/bingbot.htm} \\

\rowheight Duck.ai & 
\url{https://duckduckgo.com/duckduckgo-help-pages/results/duckduckbot} \\

\rowheight ERNIE & 
\url{https://www.baidu.com/search/spider.html} \\

\rowheight Gemini & 
\url{https://developers.google.com/search/docs/crawling-indexing/googlebot} \\

\rowheight Granite & 
\url{https://github.com/ibm-granite-community/granite-playground-agents?tab=readme-ov-file} \\

\rowheight Llama & 
\url{https://developers.facebook.com/docs/sharing/webmasters/crawler} \\

\rowheight Nova & 
\url{https://developer.amazon.com/support/amazonbot} \\

\rowheight Perplexity & 
\url{https://perplexity.ai/perplexitybot} \\

\rowheight Vibe & 
\url{https://docs.mistral.ai/robots} \\

\bottomrule
\end{tabular*} 
\caption{Official or publicly documented sources describing scraping agents / User-Agents for AI chatbots.}
\label{tab:llm_sources}
\end{table*}

%% file: tables/condition1_appendix_new.tex
\begin{table*}[t]
\centering
\setlength{\fboxsep}{1.3pt}
\resizebox{0.95\textwidth}{!}{
\begin{tabular*}{\textwidth}
{@{\extracolsep{\fill}} 
p{1.2cm} p{2.75cm} p{2.75cm} p{2.75cm} p{2cm} p{2cm}p{2cm}}
\midrule
\makecell[l]{\textbf{} \\ \\ \textbf{AI System}} & \makecell[l]{\\ \textbf{Self-Declared} \\ \textbf{User-Agent}} & \makecell[l]{\\ \textbf{User-Agent} \\ \textbf{Measured}}& \makecell[l]{\textbf{Full User-Agent} \\ \textbf{String}} & \makecell[l]{\textbf{Canary Token} \\\textbf{Count}}& \makecell[l]{\textbf{} \\ \textbf{Websites} \\\textbf{Visited}} \\ 
\midrule

\rowheight ChatGPT & 
\begin{tabular}[t]{@{}l@{}}
OAI-SearchBot\\ChatGPT-User \\ GPTBot
\end{tabular} &
\begin{tabular}[t]{@{}l@{}}
\colorbox{eno!40!white}{\makebox[3.0cm][l]{OAI-SearchBot}} \\ - \\ -
\end{tabular} &
\begin{tabular}[t]{@{}l@{}}
10 \\ - \\ -
\end{tabular}&
\begin{tabular}[t]{@{}l@{}}
68 \\ - \\ -
\end{tabular}&
\begin{tabular}[t]{@{}l@{}}
11 \\ - \\ -
\end{tabular}  \\
\hline
\rowheight Claude & 
\begin{tabular}[t]{@{}l@{}}
Claude-SearchBot\\
Claude-User\\
ClaudeBot\\-
\end{tabular} & 
\begin{tabular}[t]{@{}l@{}}
-\\-\\-\\\colorbox{dandelion!50!white}{\makebox[3.0cm][l]{Bravebot}}
\end{tabular} &  
\begin{tabular}[t]{@{}l@{}}
-\\-\\-\\38
\end{tabular} &
\begin{tabular}[t]{@{}l@{}}
- \\ - \\ - \\ 88
\end{tabular}&
\begin{tabular}[t]{@{}l@{}}
- \\ - \\ - \\ 20
\end{tabular}  \\
\hline
\rowheight Copilot & 
\begin{tabular}[t]{@{}l@{}}
Bingbot\\MicrosoftPreview
\end{tabular} & 
\begin{tabular}[t]{@{}l@{}}
\colorbox{eno!40!white}{\makebox[3.0cm][l]{Bingbot}} \\ -
\end{tabular} & 
\begin{tabular}[t]{@{}l@{}}
37 \\ -
\end{tabular}&\begin{tabular}[t]{@{}l@{}}
57 \\ -
\end{tabular}&\begin{tabular}[t]{@{}l@{}}
17 \\ -
\end{tabular}  \\ 
\hline
\rowheight Duck.ai & 
\begin{tabular}[t]{@{}l@{}}
DuckAssistBot\\
DuckDuckBot
\end{tabular} & 
\begin{tabular}[t]{@{}l@{}}
\colorbox{eno!40!white}{\makebox[3.0cm][l]{DuckAssistBot}}\\-
\end{tabular} & 
\begin{tabular}[t]{@{}l@{}}
1\\-
\end{tabular}&\begin{tabular}[t]{@{}l@{}}
56 \\ -
\end{tabular}&\begin{tabular}[t]{@{}l@{}}
9 \\ -
\end{tabular} \\
\hline
\rowheight ERNIE & 
\begin{tabular}[t]{@{}l@{}}
Baiduspider\\
ERNIEBot \\-
\end{tabular} & 
\begin{tabular}[t]{@{}l@{}}
-\\-\\\colorbox{copper!40!white}{\makebox[3.0cm][l]{Chrome}}
\end{tabular} & \begin{tabular}[t]{@{}l@{}}
-\\-\\31
\end{tabular} &
\begin{tabular}[t]{@{}l@{}}
- \\ - \\ 6
\end{tabular}&
\begin{tabular}[t]{@{}l@{}}
- \\ - \\ 1
\end{tabular} \\
\hline
\rowheight Gemini & 
\begin{tabular}[t]{@{}l@{}}
Googlebot\\Google-Extended
\end{tabular} & 
\begin{tabular}[t]{@{}l@{}}
\colorbox{eno!40!white}{\makebox[3.0cm][l]{Googlebot}}\\-
\end{tabular} &\begin{tabular}[t]{@{}l@{}}
5, 6\\-
\end{tabular} &
\begin{tabular}[t]{@{}l@{}}
78, 43 \\ -
\end{tabular}&
\begin{tabular}[t]{@{}l@{}}
12, 6 \\ -
\end{tabular}  \\ \hline
\rowheight GLM & - & - & - & -  \\
\hline
\rowheight Granite & GranitePlayground & 
\begin{tabular}[t]{@{}l@{}}
\colorbox{eno!40!white}{\makebox[3.0cm][l]{GranitePlayground}}
\end{tabular} & 2 & 75 & 12 \\
\hline
\rowheight Grok & \begin{tabular}[t]{@{}l@{}}
-\\-
\end{tabular} & 
\begin{tabular}[t]{@{}l@{}}
\colorbox{dandelion!50!white}{\makebox[3.0cm][l]{Googlebot}}\\
\colorbox{copper!40!white}{\makebox[3.0cm][l]{Chrome}}
\end{tabular} &\begin{tabular}[t]{@{}l@{}}
5, 6\\
11
\end{tabular}&\begin{tabular}[t]{@{}l@{}}
30, 20 \\ 129
\end{tabular}&\begin{tabular}[t]{@{}l@{}}
12, 5 \\ 19
\end{tabular} \\
\hline
\rowheight Kimi  & - &  \begin{tabular}[t]{@{}l@{}}
\colorbox{copper!40!white}{\makebox[3.0cm][l]{Chrome}}\\
\colorbox{copper!40!white}{\makebox[3.0cm][l]{Edge}}\\
\colorbox{dandelion!50!white}{\makebox[3.0cm][l]{Googlebot}}\\
\colorbox{copper!40!white}{\makebox[3.0cm][l]{Obsidian}}\\
\colorbox{copper!40!white}{\makebox[3.0cm][l]{Qaxbrowser}}\\
\colorbox{copper!40!white}{\makebox[3.0cm][l]{QQBrowser}}\\
\colorbox{copper!40!white}{\makebox[3.0cm][l]{QuarkPC}}\\
\colorbox{copper!40!white}{\makebox[3.0cm][l]{SLBrowser}}\\
\colorbox{copper!40!white}{\makebox[3.0cm][l]{WindowsWechat}}
\end{tabular}& \begin{tabular}[t]{@{}l@{}}
8, 16, 30\\
7, 9, 15, 22, 23, 24, 32, 33\\
5, 6\\
27\\
35\\
28, 29\\
18\\
17, 19\\
20, 21, 34\\
\end{tabular}&\begin{tabular}[t]{@{}l@{}}
5, 9, 8\\
6, 6, 8, 8, 7, 5, 5, 8\\
11, 3\\
8\\
8\\
4, 8\\
4\\
6, 7\\
4, 7, 4\\
\end{tabular}&\begin{tabular}[t]{@{}l@{}}
1, 1, 1\\
1, 1, 1, 1, 1, 1, 1, 1\\
3, 8\\
1\\
1\\
1, 1\\
1\\
1, 1\\
1, 1, 1\\
\end{tabular} \\
\hline
\rowheight Llama & 
\begin{tabular}[t]{@{}l@{}}
meta-externalagent \\
meta-externalfetcher\\
meta-webindexer

\end{tabular} & 
\begin{tabular}[t]{@{}l@{}}
\colorbox{eno!40!white}{\makebox[3.0cm][l]{meta-externalagent}}\\ - \\
\colorbox{eno!40!white}{\makebox[3.0cm][l]{meta-webindexer}}
\end{tabular} &\begin{tabular}[t]{@{}l@{}}
3\\ - \\
4
\end{tabular} &
\begin{tabular}[t]{@{}l@{}}
6 \\ - \\ 9
\end{tabular}&
\begin{tabular}[t]{@{}l@{}}
1 \\ - \\ 1
\end{tabular}   \\
\hline
\rowheight Nova & \begin{tabular}[t]{@{}l@{}}
Amazonbot \\
Amzn-SearchBot\\
Amzn-User

\end{tabular} & 
\begin{tabular}[t]{@{}l@{}}
\colorbox{eno!40!white}{\makebox[3.0cm][l]{Amazonbot}} \\ - \\ -
\end{tabular} & \begin{tabular}[t]{@{}l@{}}
36  \\ - \\ -
\end{tabular} &
\begin{tabular}[t]{@{}l@{}}
81 \\ - \\ -
\end{tabular}  &
\begin{tabular}[t]{@{}l@{}}
15 \\ - \\ -
\end{tabular} \\
\hline
\rowheight Perplexity & 
\begin{tabular}[t]{@{}l@{}}
PerplexityBot\\ Perplexity-User \\ -
\end{tabular} & 
\begin{tabular}[t]{@{}l@{}}
\colorbox{eno!40!white}{\makebox[3.0cm][l]{PerplexityBot}}\\ - \\ \colorbox{dandelion!50!white}{\makebox[3.0cm][l]{Googlebot}}
\end{tabular} & 
\begin{tabular}[t]{@{}l@{}}
39\\ - \\ 5, 6
\end{tabular} &
\begin{tabular}[t]{@{}l@{}}
110 \\ - \\ 10, 6
\end{tabular}&\begin{tabular}[t]{@{}l@{}}
17 \\ - \\ 7, 4
\end{tabular} \\
\hline
\rowheight Qwen & \begin{tabular}[t]{@{}l@{}}
-
\end{tabular} & 
\begin{tabular}[t]{@{}l@{}}
\colorbox{dandelion!50!white}{\makebox[3.0cm][l]{Googlebot}}
\end{tabular}& 
\begin{tabular}[t]{@{}l@{}}
6
\end{tabular} &
\begin{tabular}[t]{@{}l@{}}
103
\end{tabular}&
\begin{tabular}[t]{@{}l@{}}
20 
\end{tabular} \\
\hline
\rowheight Reka & \begin{tabular}[t]{@{}l@{}}
-\\-
\end{tabular} & 
\begin{tabular}[t]{@{}l@{}}
\colorbox{dandelion!50!white}{\makebox[3.0cm][l]{Googlebot}}\\
\colorbox{copper!40!white}{\makebox[3.0cm][l]{Chrome}}
\end{tabular}& 
\begin{tabular}[t]{@{}l@{}}
5\\
12, 13
\end{tabular} &
\begin{tabular}[t]{@{}l@{}}
3 \\ 56, 52
\end{tabular}&
\begin{tabular}[t]{@{}l@{}}
3 \\ 10, 8
\end{tabular} \\
\hline
\rowheight Solar & 
\begin{tabular}[t]{@{}l@{}}
- \\ - \\ -
\end{tabular} & 
\begin{tabular}[t]{@{}l@{}}
\colorbox{copper!40!white}{\makebox[3.0cm][l]{Chrome}} \\ \colorbox{copper!40!white}{\makebox[3.0cm][l]{Safari}} \\ \colorbox{copper!40!white}{\makebox[3.0cm][l]{Firefox}}
\end{tabular} & 
\begin{tabular}[t]{@{}l@{}}
25 \\ 14 \\ 26
\end{tabular} &
\begin{tabular}[t]{@{}l@{}}
29 \\ 2 \\ 2
\end{tabular}&
\begin{tabular}[t]{@{}l@{}}
11 \\ 1 \\ 1
\end{tabular} \\
\hline
\rowheight Venice & - & 
\begin{tabular}[t]{@{}l@{}}
\colorbox{dandelion!50!white}{\makebox[3.0cm][l]{Bravebot}}
\end{tabular} &38&
63 & 17 \\
\hline
\rowheight Vibe & 
\begin{tabular}[t]{@{}l@{}}
MistralAI-User\\ MistralAI-Index \\
-
\end{tabular}
& 
\begin{tabular}[t]{@{}l@{}}
-\\-\\\colorbox{dandelion!50!white}{\makebox[3.0cm][l]{Bravebot}}
\end{tabular} & 
\begin{tabular}[t]{@{}l@{}}
-\\-\\38
\end{tabular} & 
\begin{tabular}[t]{@{}l@{}}
- \\ - \\ 66
\end{tabular}& 
\begin{tabular}[t]{@{}l@{}}
- \\ - \\ 19
\end{tabular} \\

\bottomrule
\end{tabular*} 

}
\vspace{0.1cm}
\caption{Full Stage 1 Results. Measured User-Agents and references to their index within Table 10 below and compared against officially declared User-Agents (if available, see Table~\ref{tab:llm_sources}). Number of tokens measured and number of websites visited (out of 20) are shown for each match.}
\label{tab:main_results_fullstring}
\end{table*}

%% file: tables/all-uas.tex
\clearpage
\onecolumn

{\small
\begin{longtable}{| p{.025\textwidth} | p{.975\textwidth} |} 

    \caption{Full List of User-Agent Strings Measured During Stage 1\label{tab:ua-strings}}\\
    
    \hline
    \textbf{\#} & \textbf{Full User-Agent String Measured} \\ 
    \hline
    \endfirsthead
    \hline
    \textbf{\#} & \textbf{Full User-Agent String Measured} \\ 
    \hline
    \endhead
    
    1  & DuckAssistBot/1.2; (+http://duckduckgo.com/duckassistbot.html)                                                                                                                                                                                    \\
2  & GranitePlayground/1.0 (https://www.ibm.com/granite/playground; RTE@ibm.com)                                                                                                                                                                       \\
3  & meta-externalagent/1.1 (+https://developers.facebook.com/docs/sharing/webmasters/crawler)                                                                                                                                                         \\
4  & meta-webindexer/1.1 (+https://developers.facebook.com/docs/sharing/webmasters/crawler)                                                                                                                                                            \\
5  & Mozilla/5.0 (Linux; Android 6.0.1; Nexus 5X Build/MMB29P) AppleWebKit/537.36 (KHTML, like Gecko) Chrome/144.0.7559.132 Mobile Safari/537.36 (compatible; Googlebot/2.1; +http://www.google.com/bot.html)                                          \\
6  & Mozilla/5.0 (Linux; Android 6.0.1; Nexus 5X Build/MMB29P) AppleWebKit/537.36 (KHTML, like Gecko) Chrome/145.0.7632.116 Mobile Safari/537.36 (compatible; Googlebot/2.1; +http://www.google.com/bot.html)                                          \\
7  & Mozilla/5.0 (Macintosh; Intel Mac OS X 10\_15\_7) AppleWebKit/537.36 (KHTML, like Gecko) Chrome/114.0.0.0 Safari/537.36 Edg/114.0.1823.82                                                                                                         \\
8  & Mozilla/5.0 (Macintosh; Intel Mac OS X 10\_15\_7) AppleWebKit/537.36 (KHTML, like Gecko) Chrome/128.0.0.0 Safari/537.36                                                                                                                           \\
9  & Mozilla/5.0 (Macintosh; Intel Mac OS X 10\_15\_7) AppleWebKit/537.36 (KHTML, like Gecko) Chrome/130.0.0.0 Safari/537.36 Edg/130.0.0.0                                                                                                             \\
10 & Mozilla/5.0 (Macintosh; Intel Mac OS X 10\_15\_7) AppleWebKit/537.36 (KHTML, like Gecko) Chrome/131.0.0.0 Safari/537.36; compatible; OAI-SearchBot/1.3; +https://openai.com/searchbot                                                             \\
11 & Mozilla/5.0 (Macintosh; Intel Mac OS X 10\_15\_7) AppleWebKit/537.36 (KHTML, like Gecko) Chrome/139.0.0.0 Safari/537.36                                                                                                                           \\
12 & Mozilla/5.0 (Macintosh; Intel Mac OS X 10\_15\_7) AppleWebKit/537.36 (KHTML, like Gecko) Chrome/145.0.0.0 Safari/537.36                                                                                                                           \\
13 & Mozilla/5.0 (Macintosh; Intel Mac OS X 10\_15\_7) AppleWebKit/537.36 (KHTML, like Gecko) Chrome/146.0.0.0 Safari/537.36                                                                                                                           \\
14 & Mozilla/5.0 (Macintosh; Intel Mac OS X 13\_4\_1) AppleWebKit/605.1.15 (KHTML, like Gecko) Version/16.5 Safari/605.1.15                                                                                                                            \\
15 & Mozilla/5.0 (Windows NT 10.0; Win64; x64) AppleWebKit/537.36 (KHTML, like Gecko) Chrome/107.0.0.0 Safari/537.36 Edg/107.0.1418.24                                                                                                                 \\
16 & Mozilla/5.0 (Windows NT 10.0; Win64; x64) AppleWebKit/537.36 (KHTML, like Gecko) Chrome/109.0.0.0 Safari/537.36                                                                                                                                   \\
17 & Mozilla/5.0 (Windows NT 10.0; Win64; x64) AppleWebKit/537.36 (KHTML, like Gecko) Chrome/109.0.0.0 Safari/537.36 SLBrowser/9.0.0.0 SLBChan/115                                                                                                     \\
18 & Mozilla/5.0 (Windows NT 10.0; Win64; x64) AppleWebKit/537.36 (KHTML, like Gecko) Chrome/112.0.0.0 Safari/537.36 QuarkPC/2.0.5.221                                                                                                                 \\
19 & Mozilla/5.0 (Windows NT 10.0; Win64; x64) AppleWebKit/537.36 (KHTML, like Gecko) Chrome/124.0.0.0 Safari/537.36 SLBrowser/9.0.5.12181 SLBChan/103 SLBVPV/64-bit                                                                                   \\
20 & Mozilla/5.0 (Windows NT 10.0; Win64; x64) AppleWebKit/537.36 (KHTML, like Gecko) Chrome/126.0.0.0 Safari/537.36 MicroMessenger/7.0.20.1781(0x6700143B) NetType/WIFI MiniProgramEnv/Windows WindowsWechat/WMPF WindowsWechat(0x63090c11)XWEB/13283 \\
21 & Mozilla/5.0 (Windows NT 10.0; Win64; x64) AppleWebKit/537.36 (KHTML, like Gecko) Chrome/126.0.0.0 Safari/537.36 MicroMessenger/7.0.20.1781(0x6700143B) NetType/WIFI MiniProgramEnv/Windows WindowsWechat/WMPF WindowsWechat(0x63090c13)XWEB/13227 \\
22 & Mozilla/5.0 (Windows NT 10.0; Win64; x64) AppleWebKit/537.36 (KHTML, like Gecko) Chrome/128.0.1 Safari/537.36 Edg/128.0.1                                                                                                                         \\
23 & Mozilla/5.0 (Windows NT 10.0; Win64; x64) AppleWebKit/537.36 (KHTML, like Gecko) Chrome/130.0.1 Safari/537.36 Edg/130.0.1                                                                                                                         \\
24 & Mozilla/5.0 (Windows NT 10.0; Win64; x64) AppleWebKit/537.36 (KHTML, like Gecko) Chrome/130.0.4896.75 Safari/537.36 Edg/130.0.1185.36                                                                                                             \\
25 & Mozilla/5.0 (Windows NT 10.0; Win64; x64) AppleWebKit/537.36 (KHTML, like Gecko) Chrome/138.0.0.0 Safari/537.36                                                                                                                                   \\
26 & Mozilla/5.0 (Windows NT 10.0; Win64; x64) AppleWebKit/537.36 (KHTML, like Gecko) Firefox/115.0                                                                                                                                                    \\
27 & Mozilla/5.0 (Windows NT 10.0; Win64; x64) AppleWebKit/537.36 (KHTML, like Gecko) obsidian/1.7.7 Chrome/128.0.6613.186 Electron/32.2.5 Safari/537.36                                                                                               \\
28 & Mozilla/5.0 (Windows NT 10.0; WOW64) AppleWebKit/537.36 (KHTML, like Gecko) Chrome/116.0.5845.97 Safari/537.36 Core/1.116.460.400 QQBrowser/13.3.6166.400                                                                                         \\
29 & Mozilla/5.0 (Windows NT 10.0; WOW64) AppleWebKit/537.36 (KHTML, like Gecko) Chrome/116.0.5845.97 Safari/537.36 Core/1.116.477.400 QQBrowser/13.5.6280.400                                                                                         \\
30 & Mozilla/5.0 (Windows NT 10.0; WOW64) AppleWebKit/537.36 (KHTML, like Gecko) Chrome/127.0.0.0 Safari/537.36                                                                                                                                        \\
31 & Mozilla/5.0 (Windows NT 10.0; WOW64) AppleWebKit/537.36 (KHTML, like Gecko) Chrome/48.0.2564.116 Safari/537.36                                                                                                                                    \\
32 & Mozilla/5.0 (Windows NT 11.0; Win64; x64) AppleWebKit/537.36 (KHTML, like Gecko) Chrome/128.1.0 Safari/537.36 Edg/128.1.0                                                                                                                         \\
33 & Mozilla/5.0 (Windows NT 11.0; Win64; x64) AppleWebKit/537.36 (KHTML, like Gecko) Chrome/134.0.0 Safari/537.36 Edg/134.0.0                                                                                                                         \\
34 & Mozilla/5.0 (Windows NT 6.1; Win64; x64) AppleWebKit/537.36 (KHTML, like Gecko) Chrome/107.0.0.0 Safari/537.36 MicroMessenger/7.0.20.1781(0x6700143B) NetType/WIFI MiniProgramEnv/Windows WindowsWechat/WMPF WindowsWechat(0x63090819) XWEB/8555  \\
35 & Mozilla/5.0 (X11; Linux aarch64) AppleWebKit/537.36 (KHTML, like Gecko) Chrome/102.0.5005.200 Safari/537.36 Qaxbrowser                                                                                                                            \\
36 & Mozilla/5.0 AppleWebKit/537.36 (KHTML, like Gecko; compatible; Amazonbot/0.1; +https://developer.amazon.com/support/amazonbot) Chrome/119.0.6045.214 Safari/537.36                                                                                \\
37 & Mozilla/5.0 AppleWebKit/537.36 (KHTML, like Gecko; compatible; bingbot/2.0; +http://www.bing.com/bingbot.htm) Chrome/116.0.1938.76 Safari/537.36                                                                                                  \\
38 & Mozilla/5.0 AppleWebKit/537.36 (KHTML, like Gecko; compatible; Bravebot/1.0; +https://search.brave.com/help/brave-search-crawler) Chrome/W.X.Y.Z Safari/537.36                                                                                    \\
39 & Mozilla/5.0 AppleWebKit/537.36 (KHTML, like Gecko; compatible; PerplexityBot/1.0; +https://perplexity.ai/perplexitybot) \\

\bottomrule

\end{longtable}
}
\clearpage
\twocolumn

%% file: MyCite.bib
@misc{DarkVisitors,
  title={Agents},
  url={https://darkvisitors.com}, 
  author={Dark Visitors}, 
  year={2025}, publisher={Bit Flip LLC}
}

@inproceedings{liu2024somesite,
  author = {Liu, Enze and Luo, Elisa and Shan, Shawn and Voelker, Geoffrey M. and Zhao, Ben Y. and Savage, Stefan},
  title = {Somesite I Used To Crawl: Awareness, Agency and Efficacy in Protecting Content Creators From AI Crawlers},
  year = {2025},
  isbn = {9798400718601},
  booktitle = {Proceedings of the ACM Internet Measurement Conference (IMC)}
}

@inproceedings{lewis2020retrieval,
  title={{Retrieval-augmented generation for knowledge-intensive NLP tasks}},
  author={Lewis, Patrick and Perez, Ethan and Piktus, Aleksandra and others},
  booktitle={{Proc. of NeurIPs}},
  year={2020}
}

@misc{Davis_2023,
  title={ChatGPT can now search the web in Real time},
  url={https://www.theverge.com/2023/9/27/23892781/openai-chatgpt-live-web-results-browse-with-bing},
  journal={The Verge},
  publisher={Vox Media},
  author={Davis, Wes},
  year={2023},
  month={Sep}
}

@article{zychlinski2025newworldcreatingparallelpoisoned,
  title={A whole new world: Creating a parallel-poisoned web only ai-agents can see},
  author={Zychlinski, Shaked},
  journal={arXiv preprint arXiv:2509.00124},
  year={2025}
}

@inproceedings{10.1145/1242572.1242726,
  title={A large-scale study of robots. txt},
  author={Sun, Yang and Zhuang, Ziming and Giles, C Lee},
  booktitle={Proc. of the ACM international conference on World Wide Web},
  year={2007}
}

@inproceedings{brach2025ghosts,
  title={Ghosts in the Markup: Techniques to Fight Large Language Model-Powered Web Scrapers},
  author={Brach, William and Petrik, Matej and Ko{\v{s}}t’{\'a}l, Kristi{\'a}n and Ries, Michal},
  booktitle={IEEE Open Innovations Association},
  year={2025}
}

@article{raffel2020t5,
  author  = {Raffel, Colin and Shazeer, Noam and Roberts, Adam and Lee, Katherine and
       Narang, Sharan and Matena, Michael and Zhou, Yanqi and Li, Wei and Liu, Peter J.},
  title   = {Exploring the Limits of Transfer Learning with a Unified Text-to-Text Transformer},
  journal = {Journal of Machine Learning Research},
  year  = {2020}
}

@article{radford2019gpt2,
  title={Language models are unsupervised multitask learners},
  author={Radford, Alec and Wu, Jeffrey and Child, Rewon and Luan, David and Amodei, Dario and Sutskever, Ilya and others},
  year = {2019}
}

@misc{faker,
  author = {Daniele Faraglia},
  title = {faker},
  year = {2025},
  publisher = {GitHub},
  journal = {GitHub repository},
  url = {https://github.com/joke2k/faker}
}

@article{rao2025detecting,
  title={Detecting LLM-generated peer reviews},
  author={Rao, Vishisht Srihari and Kumar, Aounon and Lakkaraju, Himabindu and Shah, Nihar B},
  journal={PLoS One},
  year={2025},
  publisher={Public Library of Science San Francisco, CA USA}
}

@misc{vessum_google_index,
author = {Vessum, Steven},
  title = {How Long Does It Take for {Google} to Index a Website?},
  url = {https://www.conductor.com/academy/google-index/faq/indexing-speed/},
  year = {2025}
}

@misc{ridings_2022,
author = {Ridings, Liam},
  title = {How Long Will It Take {Google} to Index My Site?},
  url = {https://www.safaridigital.com.au/blog/how-long-will-it-take-google-to-index-my-site/},
  month   = jan,
  year    = {2022}
}

@article{zhang2015detecting,
  title={Detecting malicious activities with user-agent-based profiles},
  author={Zhang, Yang and Mekky, Hesham and Zhang, Zhi-Li and Torres, Ruben and Lee, Sung-Ju and Tongaonkar, Alok and Mellia, Marco},
  journal={International Journal of Network Management},
  year={2015},
  publisher={Wiley Online Library}
}

@article{math13193153,
  title={Weighted Transformer Classifier for User-Agent Progression Modeling, Bot Contamination Detection, and Traffic Trust Scoring},
  author={Lucz, Geza and Forstner, Bertalan},
  journal={MDPI Mathematics},
  year={2025}
}

@article{tanaka2020bot,
  title={Bot detection model using user agent and user behavior for web log analysis},
  author={Tanaka, Takamasa and Niibori, Hidekazu and Nomura, Shimpei and Kawashima, Hiroki and Tsuda, Kazuhiko and others},
  journal={Procedia Computer Science},
  year={2020},
  publisher={Elsevier}
}

@misc{Belanger_2025,
   title={Lawsuit: Reddit caught Perplexity “red-handed” stealing data from Google results},
   url={https://arstechnica.com/tech-policy/2025/10/reddit-sues-to-block-perplexity-from-scraping-google-search-results/},
   journal={Ars Technica},
   author={Belanger, Ashley},
   year={2025},
   month=oct,
   language={en}
}

@misc{spoofing_news, 
   title={The Great Masquerade: How AI Agents Are Spoofing Their Way In},
   url={https://datadome.co/agent-trust-management/ai-agent-spoofing/},
   journal={DataDome}, 
   author={Segura, Jerome}, 
   language={en-US},
   year = {2025}
 }

@inproceedings{li2021good,
  title={Good bot, bad bot: Characterizing automated browsing activity},
  author={Li, Xigao and Azad, Babak Amin and Rahmati, Amir and Nikiforakis, Nick},
  booktitle={Proc. of IEEE Symposium on Security and Privacy (SP)},
  year={2021}
}

@misc{openaibots,
  title    = {OpenAI Platform},
  author     = {OpenAI},
  howpublished = {\url{https://platform.openai.com/docs/bots}},
  year     = {2025}
}

@misc{openrouter,
  title    = {Models},
  author     = {OpenRouter},
  howpublished = {\url{https://openrouter.ai/models?order=top-weekly}},
  year     = {2025}
}

@misc{lmarena,
  title    = {Text arena | LMArena},
  author     = {LMArena},
  howpublished = {\url{https://lmarena.ai/leaderboard/text}},
  year     = {2025}
}

@article{ji_survey_2023,
  title={Survey of hallucination in natural language generation},
  author={Ji, Ziwei and Lee, Nayeon and Frieske, Rita and Yu, Tiezheng and Su, Dan and Xu, Yan and Ishii, Etsuko and Bang, Ye Jin and Madotto, Andrea and Fung, Pascale},
  journal={ACM Computing Surveys},
  year={2023}
}

@inproceedings{shan_nightshade_2024,
  title={Nightshade: Prompt-Specific Poisoning Attacks on Text-to-Image Generative Models},
  author={Shan, Shawn and Ding, Wenxin and Passananti, Josephine and Wu, Stanley and Zheng, Haitao and Zhao, Ben Y},
  booktitle={Proc. of IEEE Symposium on Security and Privacy},
  year={2024}
}

@inproceedings{cui2025odyssey,
  title={The Odyssey of robots. txt Governance: Measuring Convention Implications of Web Bots in Large Language Model Services},
  author={Cui, Jian and Zha, Mingming and Wang, XiaoFeng and Liao, Xiaojing},
  booktitle={Proc. of the ACM Computer and Communications Security (CCS)},
  year={2025}
}

@inproceedings{steinacker2026misinformation,
  title={Is Misinformation More Open? A Study of robots.txt Gatekeeping on the Web},
  author={Steinacker-Olsztyn, Nicolas and Gosain, Devashish and Dao, Ha},
  booktitle={Proc. of the ACM Web Conference},
  year={2026}
}

@article{Zhang2025RethinkingWC,
  title={Rethinking Web Cache Design for the AI Era},
  author={Yazhuo Zhang and Jin Cai and Avani Wildani and Ana Klimovic},
  journal={Proc. of the ACM Symposium on Cloud Computing},
  year={2025},
  url={https://api.semanticscholar.org/CorpusID:284593848}
}

@inproceedings{zhong2025web,
  title={Web Intellectual Property at Risk: Preventing Unauthorized Real-Time Retrieval by Large Language Models},
  author={Zhong, Yisheng and Wen, Yizhu and Guo, Junfeng and Kafai, Mehran and Huang, Heng and Guo, Hanqing and Zhu, Zhuangdi},
  booktitle={Proc. of the 2025 Conference on Empirical Methods in Natural Language Processing},
  year={2025}
}

@article{borysenko2026developer,
  title={Developer Experience with AI Coding Agents: HTTP Behavioral Signatures in Documentation Portals},
  author={Borysenko, Oleksii},
  journal={arXiv:2604.02544},
  year={2026}
}

@inproceedings{lee2026data,
  title={How do data owners say no? A case study of data consent mechanisms in web-scraped vision-language AI training datasets},
  author={Lee, Chung Peng and Hong, Rachel and Jiang, Harry H and Plotnik, Aster and Agnew, William and Morgenstern, Jamie Heather},
  booktitle={Proc. of the AAAI Conference on Artificial Intelligence},
  year={2026}
}

@inproceedings{dinzinger2024longitudinal,
  title={A Longitudinal Study of Content Control Mechanisms},
  author={Dinzinger, Michael and Granitzer, Michael},
  booktitle={Proc. of the ACM Web Conference},
  year={2024}
}

@article{dinzinger2024survey,
  title={A survey of web content control for generative AI},
  author={Dinzinger, Michael and He{\ss}, Florian and Granitzer, Michael},
  journal={arXiv preprint arXiv:2404.02309},
  year={2024}
}

@article{nakano2022webgptbrowserassistedquestionansweringhuman,
  title={Webgpt: Browser-assisted question-answering with human feedback},
  author={Nakano, Reiichiro and Hilton, Jacob and Balaji, Suchir and Wu, Jeff and Ouyang, Long and Kim, Christina and Hesse, Christopher and Jain, Shantanu and Kosaraju, Vineet and Saunders, William and others},
  journal={arXiv:2112.09332},
  year={2021}
}

@misc{pplx_online, 
 title={The first-of-its-kind Online LLM API}, 
 url={https://www.perplexity.ai/hub/blog/introducing-pplx-online-llms},  
 journal={Perplexity}, 
 author={Perplexity Team}, 
 year={2023}, month=nov, language={en}
 }

@misc{Claude_web_search_explained_2025, url={https://www.tryprofound.com/blog/what-is-claude-web-search-explained}, journal={Profound}, year={2025}, month=mar, language={en}, author={Blyskal, Josh}}

@misc{chatgpt-search,  title={Introducing ChatGPT search},
  url = {https://openai.com/index/introducing-chatgpt-search/}, journal={OpenAI}, year={2024}, month=oct, language={en}, author={OpenAI}
}

@misc{fielding2022rfc,
  title={RFC 9110: Http Semantics},
  author={Fielding, Roy and Nottingham, Mark and Reschke, Julian},
  year={2022},
  publisher={RFC Editor}
}

@article{khder2021web,
  title={Web scraping or web crawling: State of art, techniques, approaches and application.},
  author={Khder, Moaiad Ahmad},
  journal={International Journal of Advances in Soft Computing \& Its Applications},
  year={2021}
}

@misc{kimi_web_search,
  author={Kimi},
  title = {Use {Kimi} {API}'s {Internet} {Search} {Functionality}},
  note = {\url{https://platform.kimi.ai/docs/guide/use-web-search}},
  year={2026}
}

@misc{mistral_web_search,
  author={Mistral AI},
  title = {The all new le {Chat}, {Your} {AI} assistant for life and work {Mistral} {AI}},
  note = {\url{https://mistral.ai/news/all-new-le-chat}},
  year={2026}
}

@misc{grok_web_search,
  author={xAI},
  title = {{xAI} {Creators} of {Grok}, the {AI} {Chatbot}},
  note = {\url{https://x.ai/news/grok-4}},
  year={2026}
}

@misc{bain_ai_search,
  author={Bain},
  title = {Consumer reliance on {AI} search results signals new era of marketing},
  url = {https://www.bain.com/about/media-center/press-releases/20252/consumer-reliance-on-ai-search-results-signals-new-era-of-marketing--bain--company-about-80-of-search-users-rely-on-ai-summaries-at-least-40-of-the-time-on-traditional-search-engines-about-60-of-searches-now-end-without-the-user-progressing-to-a/},
  language = {en},
  year = {2026}
}

@misc{mckinsey_ai_search,
  author = {McKinsey},
  title = {Winning in the age of {AI} search},
  url = {https://www.mckinsey.com/capabilities/growth-marketing-and-sales/our-insights/new-front-door-to-the-internet-winning-in-the-age-of-ai-search},
  year={2026}
}

@misc{sommerfeld_goodbye_2025,
  title = {Goodbye {Clicks}, {Hello} {AI}: {Zero}-{Click} {Search} {Redefines} {Marketing}},
  shorttitle = {Goodbye {Clicks}, {Hello} {AI}},
  url = {https://www.bain.com/insights/goodbye-clicks-hello-ai-zero-click-search-redefines-marketing/},
  language = {en},
  urldate = {2026-04-27},
  journal = {Bain},
  author = {Sommerfeld, Natasha and McCurry, Megan and Harrington, Doug},
  month = feb,
  year = {2025}
}

@misc{malik_chatgpt_2026,
  title = {{ChatGPT} reaches {900M} weekly active users},
  url = {https://techcrunch.com/2026/02/27/chatgpt-reaches-900m-weekly-active-users/},
  language = {en-US},
  urldate = {2026-04-27},
  journal = {TechCrunch},
  author = {Malik, Aisha},
  month = feb,
  year = {2026},
}

@misc{perez_claudes_2026,
  title = {Claude's consumer growth surge continues after {Pentagon} deal debacle},
  url = {https://techcrunch.com/2026/03/06/claudes-consumer-growth-surge-continues-after-pentagon-deal-debacle/},
  language = {en-US},
  urldate = {2026-04-27},
  journal = {TechCrunch},
  author = {Perez, Sarah},
  month = mar,
  year = {2026},
}

@techreport{koster_robots_2022,
  type = {Request for {Comments}},
  title = {Robots {Exclusion} {Protocol}},
  url = {https://datatracker.ietf.org/doc/rfc9309},
  urldate = {2026-04-06},
  institution = {Internet Engineering Task Force},
  author = {Koster, Martijn and Illyes, Gary and Zeller, Henner and Sassman, Lizzi},
  month = sep,
  year = {2022}
}

@misc{OAIBotDescription:online,
author = {OpenAI},
  title = {Overview of {OpenAI} Crawlers},
  url = "https://developers.openai.com/api/docs/bots",
  month = apr,
  year = {2026}
}

@misc{venice_search,
author = {Venice.ai},
  title = {How to Use an {AI} Research Assistant: Mastering Web Search \& Document Analysis in Venice},
  url = {https://venice.ai/blog/how-to-use-anai-research-assistant-mastering-web-search-document-analysis-in-venice},
  month = feb,
  year = {2025}
}

@misc{stepfun_search,
author = {StepFun},
  title = {Step 3.5 Flash},
  url = {https://static.stepfun.com/blog/step-3.5-flash/},
  month = {feb},
  year = {2026}
}

@misc{solar_search,
  author = {Upstage PR Team},
  title = {Introducing Solar Pro 2},
  url = {https://www.upstage.ai/news/solar-pro-2},
  month = jul,
  year = {2025}
}

@misc{reka_search,
  author = {Reka},
  title = {Web Search Configuration},
  url = {https://docs.reka.ai/research/web-search},
  year = {2026}
}

@misc{qwen_search,
  author = {Qwen},
  title = {Qwen},
  url = {https://qwen.ai/home},
  month = mar,
  year = {2025}
}

@misc{amazon_search,
  author = {Amazon},
  title = {Web Grounding},
  url = {https://docs.aws.amazon.com/nova/latest/nova2-userguide/web-grounding.html},
  year = {2026}
}

@misc{llama_search,
  author = {Meta},
  title = {Llama 3.1},
  url = {https://www.llama.com/docs/model-cards-and-prompt-formats/llama3_1/},
  year = {2026}
}

@misc{liquid_search,
  author = {Liquid AI},
  title = {Liquid Playground},
  url = {https://playground.liquid.ai/chat},
  year = {2026}
}

@misc{hunyuan_search,
  author = {Baidu},
  title = {Tencent HY},
  url = {https://baike.baidu.com/en/item/Tencent%20HY/1450766},
  year = {2026}
}

@misc{granite_search,
  author = {IBM},
  title = {IBM Granite Playground | See How It Works},
  url = {https://www.ibm.com/granite/playground},
  year = {2026}
}

@misc{zai_search,
  author = {Z.AI},
  title = {Web Search},
  url = {https://docs.z.ai/guides/tools/web-search},
  year = {2026}
}

@misc{gemini_search,
  author = {Google},
  title = {Grounding with Google Search},
  url = {https://ai.google.dev/gemini-api/docs/google-search},
  month = apr,
  year = {2026}
}

@misc{ernie_search,
  author = {Baidu},
  title = {ERNIE Bot},
  url = {https://baike.baidu.com/en/item/ERNIE%20Bot/16840},
  year = {2026}
}

@misc{duck_search,
  author = {DuckDuckGo},
  title = {Duck.ai},
  url = {https://duckduckgo.com/duckduckgo-help-pages/duckai},
  year = {2026}
}

@misc{deepseek_search,
  author = {DeepSeek},
  title = {DeepSeek-R1},
  url = {https://github.com/deepseek-ai/DeepSeek-R1},
  month = apr,
  year = {2025}
}

@misc{copilot_search,
  author = {Microsoft},
  title = {Copilot Search in Bing},
  url = {https://www.microsoft.com/en-us/bing/copilot-search/?form=MA13XW},
  year = {2026}
}

@misc{ClaudeUseBing,
  author = {Kyle Wiggers},
  title = {Anthropic appears to be using Brave to power web search for its Claude chatbot},
  url = "https://techcrunch.com/2025/03/21/anthropic-appears-to-be-using-brave-to-power-web-searches-for-its-claude-chatbot/",
  month = {3},
  year = {2025}
}

@misc{nytimes_v_openai_2023,
  title={The New York Times Company v. OpenAI, Inc. and Microsoft Corporation},
  author={{The New York Times Company}},
  year={2023},
  month={December},
  day={27},
  howpublished={Complaint filed in U.S. District Court for the Southern District of New York},
  note={Case No. 1:23-cv-11195 (S.D.N.Y.). Lead case in consolidated litigation. Motion to dismiss denied March 2025}
}

@misc{silverman_v_meta_2023,
  title={Silverman v. Meta Platforms, Inc.},
  author={{Silverman, Sarah} and {Golden, Christopher} and {Kadrey, Richard}},
  year={2023},
  month={July},
  day={7},
  howpublished={Class Action Complaint filed in U.S. District Court for the Northern District of California},
  note={Case No. 3:23-cv-03417 (N.D. Cal.). Companion case to Silverman v. OpenAI}
}

@misc{bartz_v_anthropic_2024,
  title={Bartz v. Anthropic PBC},
  author={{Bartz, Andrea} and {Graeber, Charles} and {Johnson, Kirk Wallace}},
  year={2024},
  month={August},
  howpublished={Class Action Complaint filed in U.S. District Court for the Northern District of California},
  note={Case No. 3:24-cv-04546 (N.D. Cal.)}
}

@misc{intercept_v_openai_2024,
  title={The Intercept Media, Inc. v. OpenAI, Inc. and Microsoft Corporation},
  author={{The Intercept Media, Inc.}},
  year={2024},
  month={January},
  day={5},
  howpublished={Complaint filed in U.S. District Court for the Southern District of New York},
  note={Case No. 1:24-cv-00148 (S.D.N.Y.). Independent news organization copyright claims}
}

@misc{disney_v_midjourney_2025,
  title={The Walt Disney Company v. Midjourney, Inc.},
  author={ForensisGroup},
  year={2025},
  month={June},
  day={12},
  howpublished={Complaint filed in U.S. District Court}
}

@misc{kadrey_v_meta_2023,
  title={Kadrey v. Meta Platforms, Inc.},
  author={{Kadrey, Richard}},
  year={2023},
  month={July},
  day={7},
  howpublished={Class Action Complaint filed in U.S. District Court for the Northern District of California},

}

@misc{ziff_davis_v_openai_2025,
  title={Ziff Davis, Inc. v. OpenAI, Inc.},
  author={{Ziff Davis, Inc.}},
  year={2025},
  howpublished={Complaint filed in U.S. District Court},
  note={Technology media company v. OpenAI. Recent filing}
}

@misc{Perplexi-fakeUA,
author = {Lorenzo Franceschi-Bicchierai},
  title = {Perplexity accused of scraping websites that explicitly blocked AI scraping},
  url = "https://techcrunch.com/2025/08/04/perplexity-accused-of-scraping-websites-that-explicitly-blocked-ai-scraping/",
  month = {8},
  year = {2025}
}

@string{CCS= "ACM Conference on Computer and Communications Security (CCS)"}

@inproceedings{kim2025scrapers,
  title={Scrapers selectively respect robots.txt directives: evidence from a large-scale empirical study},
  author={Kim, Taein and Bock, Karstan and Luo, Claire and Liswood, Amanda and Poroslay, Chloe and Wenger, Emily},
  booktitle={Proc. of the ACM Internet Measurement Conference (IMC)},
  year={2025}
}

@inproceedings{shan2020fawkes,
  title={Fawkes: Protecting privacy against unauthorized deep learning models},
  author={Shan, Shawn and Wenger, Emily and Zhang, Jiayun and Li, Huiying and Zheng, Haitao and Zhao, Ben Y},
  booktitle={Proc. of USENIX Security},
  year={2020}
}

@inproceedings{shan2023glaze,
  title={Glaze: Protecting artists from style mimicry by text-to-image models},
  author={Shan, Shawn and Cryan, Jenna and Wenger, Emily and Zheng, Haitao and Hanocka, Rana and Zhao, Ben Y},
  booktitle={Proc. of USENIX Security},
  year={2023}
}

@misc{anubis, 
  title={Anubis Github Repository}, 
  author={Xi Iaso}, 
  year={2025}, 
  note={\url{https://github.com/TecharoHQ/anubis}}
}

@misc{cloudflare_ai_product,
  title={{Cloudflare Just Changed How AI Crawlers Scrape the Internet-at-Large; Permission-Based Approach Makes Way for A New Business Model}}, 
  author={Cloudflare},
  year={2025}, 
  note={\url{https://www.cloudflare.com/press/press-releases/2025/cloudflare-just-changed-how-ai-crawlers-scrape-the-internet-at-large/}}
}

@misc{cloudflare_bots,
  author={Cloudflare},
  year={2024},
  note={\url{https://www.cloudflare.com/application-services/products/bot-management/}}
}

@article{tarpits,
  title={AI haters build tarpits to trap and trick AI scrapers that ignore robots.txt},
  author={Ashley Belanger},
  year={2025},
  journal={Ars Technica},
  note={\url{https://arstechnica.com/tech-policy/2025/10/reddit-sues-to-block-perplexity-from-scraping-google-search-results/}}
}

@article{longpre2024consent,
  title={Consent in crisis: the rapid decline of the AI data commons},
  author={Longpre, Shayne and Robert Mahari and Ariel Lee and others},
  journal={Proc. of NeurIPS},
  year={2024}
}

@article{fayolle2026internet,
  title={On the Internet, Nobody Knows You're an LLM Bot: Unmasking Web Agents with Multi-Layer Fingerprinting},
  author={Fayolle, Iliana and Bouhenniche, Sihem and P{\'e}lissier, Samuel and Laperdrix, Pierre and Maurice, Cl{\'e}mentine and Rudametkin, Walter},
  journal={arXiv preprint arXiv:2606.30119},
  year={2026}
}

@misc{bravewebdiscovery,
    title = {What is the {Web} {Discovery} {Project}?},
    url = {https://support.brave.app/hc/en-us/articles/4409406835469-What-is-the-Web-Discovery-Project},
    language = {en-US},
    author = {Brave},
    urldate = {2026-08-06},
    journal = {Brave Help Center},
    month = nov,
    year = {2025},
}

@incollection{zhao2022web,
  title={Web scraping},
  author={Zhao, Bo},
  booktitle={Encyclopedia of big data},
  pages={951--953},
  year={2022},
  publisher={Springer}
}

@misc{ai_amok,
    title = {Artificial intelligence web crawlers are running amok},
    url = {https://www.npr.org/2024/07/05/nx-s1-5026932/artificial-intelligence-web-crawlers-are-running-amok},
    urldate = {2026-08-11},
    journal = {NPR},
    author = {Allyn, Bobby},
    month = jul,
    year = {2024},
}

@techreport{jimenez-tbd-robotstxt-update-00,
    number =    {draft-jimenez-tbd-robotstxt-update-00},
    type =      {Internet-Draft},
    institution =   {Internet Engineering Task Force},
    publisher = {Internet Engineering Task Force},
    note =      {Work in Progress},
    url =       {https://datatracker.ietf.org/doc/draft-jimenez-tbd-robotstxt-update/00/},
    author =    {Jaime Jimenez},
    title =     {{Robots.txt update proposal}},
    pagetotal = 5,
    year =      2024,
    month =     nov,
    day =       6,
}

@article{lopez2026generative,
  title={Do Generative AI Assistants Respect robots. txt? Tracing Web Access Beyond Visible Answers},
  author={Lopez-Fonseca, Gabriel and Rodriguez, David and Bechtold, Stefan and Del Alamo, Jose M},
  journal={arXiv preprint arXiv:2607.14447},
  year={2026}
}
